\documentclass[fleqn,10pt]{wlscirep}
\usepackage[numbers]{natbib}
\pdfoutput=1
\usepackage{graphicx}
\usepackage{tabularx}
\usepackage{wrapfig}
\usepackage{etoolbox}
\usepackage{float}
\usepackage{balance}
\usepackage{pgfplotstable} 
\usepackage{pgfplots}
\pgfplotsset{compat=newest}
\usepackage{tikz}
\usepackage{hyperref}
\usepackage{subcaption} % For subfigures if needed in the future
\usetikzlibrary{shapes,arrows,positioning}
\usepgfplotslibrary{statistics} % Load the statistics library
\usepackage{pstricks}
\pgfplotsset{compat=newest} % Sets the compatibility level
\usepackage{xcolor}
\usepackage{amsmath}
\usepackage{graphicx} % Include this package to add figures
\usepackage{cleveref}
\usepackage[font={up}]{caption}

\title{Why and How do Complex Systems Self-Organize at All? Average Action Efficiency as a Predictor, Measure, Driver, and Mechanism of Self-Organization}

\author[1]{
    Matthew J Brouillet
}
\author[1,2,+]{
Georgi Yordanov Georgiev\thanks{
This preprint paper was submitted to the Journal "Nature" on 17 July 2024. The final published version may differ from this version.
This work is licensed under a Creative Commons Attribution-NonCommercial-ShareAlike 4.0 International License.
}
}

\affil[1]{
Assumption University, Worcester, MA 01609, USA}
\affil[2]{Worcester Polytechnic Institute, Worcester, MA}
\affil[+]{Corresponding author: ggeorgiev@wpi.edu, ggeorgie@assumption.edu, ORCID 0000-0002-6565-8589}
\begin{abstract}
Self-organization in complex systems is a process in which randomness is reduced and emergent structures appear that allow the system to function in a more competitive way with other states of the system or with other systems. It occurs only in the presence of energy gradients, facilitating energy transmission through the system and entropy production. Being a dynamic process, self-organization requires a dynamic measure and dynamic principles. The principles of decreasing unit action and increasing total action are two dynamic variational principles that are viable to utilize in a self-organizing system. Based on this, average action efficiency can serve as a quantitative measure of the degree of self-organization. Positive feedback loops connect this measure with all other characteristics of a complex system, providing all of them with a mechanism for exponential growth, and indicating power law relationships between each of them as confirmed by data and simulations. In this study, we apply those principles and the model to agent-based simulations. We find that those principles explain self-organization well and that the results confirm the model. By measuring action efficiency we can have a new answer to the question: "What is complexity and how complex is a system?". This work shows the explanatory and predictive power of those models, which can help understand and design better complex systems.
\end{abstract}

\begin{document}
%\flushbottom
\maketitle

\section{Introduction}

\subsection{Background and Motivation}

Self-organization is key to understanding the existence of, and the changes in all systems that lead to higher levels of complexity and perfection in development and evolution.  It is a scientific as well as a philosophical question, as our understanding deepens and our realization of the importance of the process grows. Self-organization often leads to more efficient use of resources and optimized performance, which is one measure of the degree of perfection. By degree of perfection here we mean a more organized, robust, resilient, competitive, and alive system. Because competition for resources is always the selective evolutionary pressure in systems of different natures, the more efficient systems will survive at all levels of Cosmic Evolution. 

Our goal is to contribute to the explanation of the mechanisms of self-organization that drive Cosmic Evolution from the Big Bang to the present, and into the future, and its measures\citep{sagan1980cosmos, chaisson2002cosmic, kurzweil2005singularity}. Self-organization has a universality independent of the substrate of the system - physical, chemical, biological, or social - and explains all of its structures  \citep{de2023thermodynamics, england2022self, walker2013algorithmic, walker2019new}. Establishing a universal, quantitative, absolute method to measure the organization of any system will help us understand the mechanisms of functioning and organization in general and enable us to design specific systems with the highest level of perfection \citep{georgiev2002least, georgiev2016free, georgiev2015mechanism, georgiev2017exponential, Georgiev2021Stars}. 

Previous attempts to quantify organization have used static measures, such as information \citep{Shannon1948, Jaynes1957, GellMann1995, Yockey2005, Crutchfield2003, Williams2010, Ay2013} and entropy \citep{Kolmogorov1965, Grassberger1986, Pincus1991, Costa2002, Lizier2008, Rosso2007}, but our approach offers a new, dynamic perspective. We do not use Hamilton's stationary action principle, though, we expand it and derive from it a dynamic action principle for the system studied, where average unit action for one trajectory is continuously decreasing and total action for the whole system is continuously increasing with self-organization. 

Despite its significance, the mechanisms driving self-organization remain only partially understood due to the non-linearity in the dynamics of complex systems. Existing approaches often rely on specific metrics like entropy or information which are static, and while they are valuable, they have limitations in their universality and ability to predict the most organized state of a system. In many cases, they fail to describe the dynamics of the processes that lead to the increase of order and complexity. For example, a less organized system may require more bits to be described using information, compared to the same system in a more organized, i.e. efficient state. These traditional measures often fall short in providing a comprehensive, quantitative framework that can be applied across various types of complex systems, and to describe the mechanisms that lead to higher levels of organization. This gap in understanding highlights the need for a new measure that can universally and quantitatively assess the level of organization in complex systems and the transitions between them.

The motivation for this study stems from the desire to bridge this gap by introducing a novel measure of organization based on dynamical variational principles. More specifically we use Hamilton's principle of stationary action, which is the basis of all laws of physics. In the limiting case, when the second variation of the action is positive, this makes it a true principle of least action. The principle of least action posits that the path taken by a physical system between two states is the one for which the action is minimized. Extending this principle to complex systems, we propose Average Action Efficiency as a new, dynamic measure of organization. It quantifies the level of organization and serves as a predictive tool for determining the most organized state of a system. It also correlates with all other measures of complex systems, justifying and validating its use.

Understanding the mechanisms of self-organization has profound implications across various scientific disciplines. Understanding these natural optimization processes can inspire the development of more efficient algorithms and strategies in engineering and technology. It can enhance our understanding of biological and ecological processes. It can allow us to design more efficient economic and social systems. Studying self-organization also has profound scientific and philosophical implications. It challenges traditional notions of causality and control, emphasizing the role of local interactions and feedback loops in shaping global patterns. In our model, each characteristic of a complex system is simultaneously a cause and an effect of all others. By developing a universal, quantitative measure of organization, we aim to advance  our understanding of self-organization and provide practical tools for optimizing complex systems across different fields.

\subsection{Overview of the Theoretical Framework} 
 
We use the extension of Hamilton's Principle of Stationary Action to a Principle of Dynamic Action, according to which action in self-organizing systems is changing in two ways: decreasing the average action for one event and increasing the total amount of action in the system during the process of self-organization, growth, evolution, and development. This view can lead to a deeper understanding of the fundamental principles of nature's self-organization, evolution, and development in the universe, ourselves, and our society. 

\subsection{Hamilton's principle and action efficiency}
Hamilton's principle of stationary action is the most fundamental principle in nature, from which all other physics laws are derived \citep{Maupertuis1751LAP, goldstein:mechanics}. Everything derived from it is guaranteed to be self-consistent \citep{taylor2001hidden}. Beyond classical and quantum mechanics, relativity, and electrodynamics, it has applications in statistical mechanics, thermodynamics, biology, economics, optimization, control theory, engineering, and information theory \citep{lauster2005principle, nath2022novel, bersani2021lagrangian}. We propose its application, extension, and connection to other characteristics of complex systems as part of the complex systems theory. 

In most cases in classical mechanics, Hamilton's stationary action is minimized, in some cases, it is a saddle point, and it is never maximized. The minimization of average unit action is proposed as a driving principle and the arrow of evolutionary time, and the saddle points are temporary minima that transition to lower action states with evolution. Thus, globally, on long-time scales, average action is minimized and continuously decreasing, when there are no external limitations. This turns it into a dynamic action principle for open-ended processes of self-organization evolution and development.

Our thesis is that we can complement other measures of organization and self-organization by applying a new, absolute, and universal measure based on Hamilton's principle and its extension. This measure can be related to previously used measures, such as entropy and information, as in our model for the mechanism of self-organization,  progressive development, and evolution. We demonstrate this with power-law relationships in the results.

This paper presents a derivation of a quantitative measure of action efficiency and a model in which all characteristics of a complex system reinforce each other, leading to exponential growth and power law relations between each pair of characteristics. The principle of least action is proposed as the driver of self-organization, as agents of the system follow natural laws in their motion, resulting in the most action-efficient paths. This explains why complex systems form structures and order, and continue self-organizing in their evolution and development.

Our measure of action efficiency assumes dynamical flow networks away from thermodynamic equilibrium that transport matter and energy along their flow channels and applies to such systems. The significance of our results is that they empower natural and social sciences to quantify organization and structure in an absolute, numerical, and unambiguous way. Providing a mechanism through which the least action principle and the derived measure of average action efficiency as the level of organization interact in a positive feedback loop with other characteristics of complex systems explains the existence of observed events in Cosmic Evolution. The tendency to minimize average unit action for one crossing between nodes in a complex flow network comes from the principle of least action and is proposed as the arrow of time, the main driving principle towards, and explanation of progressive development and evolution that leads to the enormous variety of systems and structures that we observe in nature and society.

\subsection{Mechanism of Self-Organization}
The research in this study demonstrates the driving principle and mechanism of self-organization and evolution in general open, complex, non-equilibrium thermodynamic systems, employing agent-based modeling. We propose that the state with the least average unit action is the attractor for all processes of self-organization and development in the universe across all systems. We measure this state through Average Action Efficiency (AAE).

We present a model for quantitatively calculating the amount of organization in a general complex system and its correlation with all other characteristics through power law relationships. We also show the cause for the progressive development and evolution which is the positive feedback loop between all characteristics of the system that leads to an exponential growth of all of them until an external limit is reached. Always, the internal organization of all complex systems in nature reflects their external environment where the flows of energy and matter come from. This model also predicts power law relationships between all characteristics. Numerous measured complexity-size scaling relationships confirm the predictions of this model \citep{Bonner2004size, carneiro1967relationship, west2017scale}.

Our work addresses a gap in complex system science by providing an absolute and quantitative measure of organization, namely AAE, based on the movement of agents and their dynamics. This measure is functional and dynamic, not relative and static as in many other metrics. We show that the amount of organization is inversely proportional to the average physical amount of action in a system. We derive the expression for organization, apply it to a simple example, and validate it with results from agent-based modeling (ABM) simulations which allow us to verify experimental data, and to vary conditions to address specific questions\citealp{gershenson2020self, sayama2015introduction}. We discuss extensions of the model for a large number of agents and state the limitations and applicability of this model in our list of assumptions. 

Measuring the level of organization in a system is crucial because it provides a long-sought criterion for evaluating and studying the mechanisms of self-organization in natural and technological systems. All those are dynamic processes, which necessitate searching for a new, dynamic measure. By measuring the amount of organization, we can analyze and design complex systems to improve our lives, in ecology, engineering, economics, and other disciplines. The level of organization corresponds to the system's robustness, which is vital for survival in case of accidents or events endangering any system's existence \citep{carlson2002complexity}.  Philosophically and mathematically, each characteristic of the system is a cause and effect of all the others, similar to auto-catalytic cycles, which is well-studied in cybernetics \citep{heylighen2001cybernetics}. 

\subsection{Negative Feedback}
Negative feedback is evident in the fact that large deviations from the power law proportionality between the characteristics are not observed or predicted. This proportionality between all characteristics at any stage of the process of self-organization is the balanced state of functioning which is usually known as a Homeostatic, or dynamical equilibrium state of the system. Complex systems function as wholes only at values of all characteristics close to this Homeostatic state. If some external influence causes large deviations even on one of the characteristics from this homeostatic value, the system functioning is compromised\citep{heylighen2001cybernetics}. 

\subsection{Unit-Total Dualism}
We find a unit-total dualism: unit quantities of the characteristics are minimized while total quantities are maximized with systems' growth. For example, the average unit action for one event, which is one edge crossing in networks, is derived from the average path length and path time, and it is minimized as calculated by the average action efficiency $\alpha$. At the same time, the total amount of action $Q$ in the whole system increases, as the system grows, which can be seen in the results from our simulation. This is an expression of the principles of decreasing average unit action and increasing total action. Similarly, unit entropy per one trajectory decreases in self-organization, as the total entropy of the system increases with its growth, expansion, and increasing number of agents. Those can be termed the principles of decreasing unit entropy and of increasing total entropy. The information for describing one event in the system, with increased efficiency and shorter paths is decreasing, while the total information in the system as it grows is increasing. They are also related by a power law relationship, which means, that one can be correlated to the other, and for one of them to change, the other must also change proportionally. 

\subsection{Unit Total Dualism Examples}
Analogous qualities are evidenced in data for real systems and appear in some cases so often that they have special names. For example, the Jevons paradox (Jevons effect) was published in 1866 by the English economist William S. Jevons \citep{jevons1866coal} . In one example, as the fuel efficiency of cars increased, the total miles traveled also increased to increase the total fuel expenditure. This is also named a "rebound effect" from increased energy efficiency \citep{berkhout2000defining}. The naming of this effect as a "paradox" shows that it is unexpected, not well studied, and sometimes considered as undesirable. In our model, it is derived mathematically as a result of the positive feedback loops of the characteristics of complex systems, which is the mechanism of its self-organization, and supported by the simulation results. It is not only unavoidable, but also necessary for the functioning, self-organization, evolution, and development of those systems. 

In economics, it is evident that with increased efficiency, the costs decrease which increases the demand, which is named the "law of demand" \citep{hildenbrand1983law}. This is another example of a size-complexity rule, whereas the efficiency increases, which in our work is a measure of complexity, the demand increases, which means that the size of the system also increases. In the 1980s the Jevons paradox was expanded to a Khazzoom–Brookes postulate, formulated by Harry Saunders in 1992 \citep{saunders1992khazzoom}, which says that it is supported by the "growth theory" which is the prevailing economic theory for long-run economic growth and technological progress. Similar relations have been observed in other areas, such as in the Downs–Thomson paradox \citep{downs2000stuck}, where increasing road efficiency increases the number of cars driving on the road. These are just a few examples that point out that this unit-total dualism has been observed for a long time in many complex systems and it was thought to be paradoxical.

\subsection{Action Principles in this Simulation, Potential Well}
In each run of this specific simulation, the average unit action has the same stationary point, which is a true minimum of the average unit action, and the shortest path between the fixed nodes is a straight line.  This is the theoretical minimum and remains the same across simulations. The closest analogy is with a particle in free fall, where it minimizes action and falls in a straight line, which is a geodesic. The difference in the simulation is that the ants have a wiggle angle and, at each step, deposit pheromone that evaporates and diffuses, therefore the difference with gravity is that the effective attractive potential is not uniform. Due to this the potential landscape changes dynamically. The shape of the walls of the potential well changes slightly with fluctuations around the average at each step.  It also changes when the number of ants is varied between runs.  

The potential well is steeper higher on its walls, and the system cannot be trapped there in local minima of the fluctuations. This is seen in the simulation as initially, the agents form longer paths that disintegrate into shorter ones. In this region away from the minimum, the action is truly always minimized, with some stochastic fluctuations. Near the bottom of the well, the slope of its wall is smaller, and local minima of the fluctuations cannot be overcome easily by the agents. Then the the system temporarily gets trapped in one of those local minima and the average unit action is a dynamical saddle point. 

The simulation shows that with fewer ants, the system is more likely to get trapped in a local minimum, resulting in a path with greater curvature and higher final average action (lower average action efficiency) compared to the theoretical minimum. With an increasing number of ants, they can explore more neighboring states, find lower local minima, and find lower average action states. Therefore, increasing the number of ants allows the system to explore more effectively neighboring paths and find shorter ones. This is evident as the average action efficiency improves when there are more ants, which can escape higher local minima and find lower action values (see  Fig. \ref{a-N}). As the number of ants (agents, system size) increases, they asymptotically find lower local minima or lower average action, improving average action efficiency, though never reaching the theoretical minimum.

In future simulations, if the distance between nodes is allowed to shrink and external obstacles are reduced, the shape of the entire potential well changes dynamically. Its minimum becomes lower, the steepness of its walls increases and the system more easily escapes local minima. However, it still does not reach the theoretical minimum, due to its fluctuations near the minimum of the well. The average action decreases, and average action efficiency increases with the lowering of this minimum, demonstrating continuous open-ended self-organization and development. This illustrates the dynamical action principle.

\subsection{ Research Questions and Hypotheses}

This study aims to answer the following research questions:
\begin{enumerate}
    \item How can a dynamical variational action principle explain the continuous self-organization, evolution, and development of complex systems?
    \item Can Average Action Efficiency (AAE) be a measure for the level of organization of complex systems?
    \item Can the proposed positive feedback model accurately predict the self-organization processes in systems?
    \item What are the relationships between various system characteristics, such as AAE, total action, order parameter, entropy, flow rate, and others?
\end{enumerate}

Our hypotheses are:
\begin{enumerate}
    \item  A dynamical variational action principle can explain the continuous self-organization, evolution and development of complex systems.
    \item AAE is a valid and reliable measure of organization that can be applied to complex systems.
    \item The model can accurately predict the most organized state based on AAE.
    \item The model can predict the power-law relationships between system characteristics that can be quantified.
\end{enumerate}

\subsection{Summary of the Specific Objectives of the Paper}

1. Define and Apply the Dynamical Action Principle: Define and apply the dynamical action principle, which extends the classical stationary action principle to dynamic, self-organizing systems, in open-ended evolution, showing that unit action decreases while total action increases during self-organization.

2. Demonstrate the Predictive Power of the Model: Build and test a model that quantitatively and numerically measures the amount of organization in a system, and predicts the most organized state as the one with the least average unit action and highest average action efficiency. Define the cases in which action is minimized, and based on that predict the most organized state of the system. The theoretical most organized state is where the edges in a network are geodesics. Due to the stochastic nature of complex systems, those states are approached asymptotically, but in their vicinity, the action is stationary due to local minima.

3. Validate a New Measure of Organization: Based on 1 and 2, develop and apply the concept of average action efficiency, rooted in the principle of least action, as a quantitative measure of organization in complex systems.

4. Explain Mechanisms of Progressive Development and Evolution: Apply a model of positive feedback between system characteristics to predict exponential growth and power-law relationships, providing a mechanism for continuous self-organization. Test it by fitting its solutions to the simulation data, and compare them to real-world data from the literature. 

5. Simulate Self-Organization Using Agent-Based Modeling: Use agent-based modeling (ABM) to simulate the behavior of an ant colony navigating between a food source and its nest to explore how self-organization emerges in a complex system.

6. Define unit-total (local-global) dualism: Investigate and define the concept of unit-total dualism, where unit quantities are minimized while total quantities are maximized as the system grows, and explain its implications as variational principles for complex systems.

7. Contribute to the Fundamental and Philosophical Understanding of Self-Organization and Causality: Enhance the theoretical understanding of self-organization in complex systems, offering a robust framework for future research and practical applications.

This research aims to provide a robust framework for understanding and quantifying self-organization in complex systems based on a dynamical principle of decreasing unit action for one edge in a complex system represented as a network. By introducing Average Action Efficiency (AAE) and developing a predictive model based on the principle of least action, we address critical gaps in existing theories and offer new insights into the dynamics of complex systems. The following sections will delve deeper into the theoretical foundations, model development, methodologies, results, and implications of our study.

\section{Building the Model: }

\subsection{Hamilton's Principle of Stationary Action for a System}

In this work, we utilize Hamilton's Principle of Stationary Action, a variational method, to study self-organization in complex systems. Stationary action is found when the first derivative is zero. When the second variation is positive, the action is a minimum. Only in this case, do we have the true least action principle. We will discuss in what situations this is the case. Hamilton's Principle of Stationary Action asserts that the evolution of a system between two states occurs along the path that makes the action functional stationary. By identifying and extremizing this functional, we can gain a deeper understanding of the dynamics and driving forces behind self-organization and describe it from first principles. 

The classical Hamilton's principle is:

\begin{equation}
\label{eq1}
\delta I(q,p,t)= \delta \int_{t_1}^{t_2} L(q(t),\dot q(t), t)\, dt = 0
\end{equation}

where $\delta$ is an infinitesimally small variation in the action integral $I$, $L$ is the Lagrangian, $q(t)$ are the generalized coordinates, $\dot q(t)$ are the time derivatives of the generalized coordinates, $p$ is the momentum, and $t$ is the time.  $t_1$ and $t_2$ are the initial and final times of the motion.

For brevity, further in the text, we will use when appropriate $L=$ $L(q(t),\dot q(t), t)$, and $I=I(q,p,t)$. 

This is the principle from which all physics and all observed equations of motion are derived. The above equation is for one object. For a complex system, there are many interacting agents. That means that we can propose that the sum of all actions of all agents is taken into account. This sum is minimized in its most action-efficient state, which we define as being the most organized. In previous papers \cite{georgiev2002least, georgiev2015mechanism, georgiev2017exponential, Georgiev2021Stars} we have stated that for an organized system we can find the natural state of that system as the one in which the variation of the sum of actions of all of the agents is zero:

\begin{equation}
\label{eq2}
\delta \sum_{i=1}^{n} I_i = \delta \sum_{i=1}^{n} \int_{t_1}^{t_2} L_i \, dt = 0
\end{equation}

where $I_i$ is the action of the $i$-th agent, $L_i$ is the Lagrangian of the $i$-th agent, and $n$ represents the number of agents in the system, $t_1$ and $t_2$ are the initial and final times of the motions.

\textbf{A network representation of a complex system. }When we represent the system as a network, we can define one edge crossing as a unit of motion, or one event in the system, for which the unit average action efficiency is defined. In this case, the sum of the actions of all agents for all of the crossings of edges per agent per unit time, which is the total number of crossings (the flow of events, $\phi$), is the total amount of action in the network, $Q$. In the most organized state of the system, the variation of the total action, $Q$, is zero, which means that it is extremised as well and for the complex system in our example this extremum is a maximum. 

\subsection{An example of true action minimization: conditions}

\begin{enumerate}
    \item The agents are free particles, not subject to any forces, so the potential energy is a constant and can be set to be zero because the origin for the potential energy can be chosen arbitrarily, therefore $V = 0$. Then, the Lagrangian $L$ of the element is equal only to the kinetic energy $T = \frac{mv^2}{2}$ of that element: 
    \begin{equation}
    L = T - V = T = \frac{mv^2}{2}
    \end{equation}
    where $m$ is the mass of the element, and $v$ is its speed.

    \item We are assuming that there is no energy dissipation in this system, so the Lagrangian of the element is a constant:
    \begin{equation}
    L = T = \frac{mv^2}{2} = \text{constant}
    \end{equation}

    \item The mass $m$ and the speed $v$ of the element are assumed to be constants.

    \item The start point and the end point of the trajectory of the element are fixed at opposite sides of a square (see Fig. A1). This produces the consequence that the action integral cannot become zero, because the endpoints cannot get infinitely close together:
    \begin{equation}
    I = \int_{t_1}^{t_2} L \, dt = \int_{t_1}^{t_2} (T - V) \, dt = \int_{t_1}^{t_2} T \, dt \neq 0
    \end{equation}

    \item  The action integral cannot become infinity, i.e., the trajectory cannot become infinitely long:
    \begin{equation}
    I = \int_{t_1}^{t_2} L \, dt = \int_{t_1}^{t_2} (T - V) \, dt = \int_{t_1}^{t_2} T \, dt \neq \infty
    \end{equation}

    \item The agents do not interact with other agents.

    \item In each configuration of the system, the actual trajectory of the element is determined as the one with the Least Action from Hamilton's Principle:
    \begin{equation}
    \delta I = \delta \int_{t_1}^{t_2} L \, dt = \delta \int_{t_1}^{t_2} (T - V) \, dt = \delta \int_{t_1}^{t_2} T \, dt = 0
    \end{equation}

    \item The medium inside the system is isotropic (it has all its properties identical in all directions). The consequence of this assumption is that the constant velocity of the element allows us to substitute the interval of time with the length of the trajectory of the element.

    \item The second variation of the action is positive, because $V=0$, and $T>0$, therefore the action is a true minimum. 
\end{enumerate}

\subsection{Building the Model}

In our model, the organization is proportional to the inverse of the average of the sum of actions of all elements \eqref{eq:Alpha}. This is the average action efficiency and we can label it with a symbol $\left\langle \alpha \right\rangle$. Here average action efficiency measures the amount of organization of the system. In a complex network, many different arrangements can correspond to the same action efficiency and therefore have the same level of organization. Thus, the average action efficiency represents the macrostate of the system. Many possible microstates of combinations of nodes, paths, and agents on the network can correspond to the same macrostate as measured by  $\left\langle \alpha \right\rangle$. This is analogous to temperature in statistical mechanics representing a macrostate corresponding to many microstates of the molecular arrangements in the gas. 

To make the action efficiency dimensionless we multiply the numerator by Planck's constant h. Now it takes the meaning that the average action efficiency is inversely proportional to the average number of action quanta for one crossing between two nodes in the system. This also provides an absolute reference point $h$ for the measure of organization.

In general,
\begin{equation}
    \left\langle \alpha \right\rangle = \frac{hnm}{\sum_{i,j=1}^{nm} I_{i,j}} 
    \label{eq:Alpha}
\end{equation}

where n is the number of agents, and m is the average number of nodes each agent crosses per unit time. If we multiply the number of agents by the number of crossings for each agent, we can define it as the flow of events in the system per unit of time, $\phi=nm$

In general:
\begin{equation}
    \left\langle \alpha \right\rangle = \frac{h\phi}{\sum_{i,j=1}^{nm} I_{i,j}} \label{eq:6}
\end{equation}

In the denominator, the sum of all actions of all agents and all crossings is defined as the total action per unit time in the system, $Q$. 

\begin{equation}
    Q={\sum_{i,j=1}^{nm} I_{i,j}} \label{eq:7}
\end{equation}

In our simulation, the average path length is equal to the average time because the speed of the agents in the simulation is set to one patch per second.

\begin{equation}
\left\langle t \right\rangle=\left\langle l \right\rangle
\label{eq:8}
\end{equation}
When the Lagrangian does not depend on time, because the speed is constant and there is no friction, as in this simulation, the kinetic energy is a constant (assumption \#2), so the action integral takes the form:
\begin{equation}
    I = \int_{t_1}^{t_2} L dt = \int_{t_1}^{t_2} T dt = T(t_2 - t_1) = T \Delta t  = L \Delta t \label{eq:9}
\end{equation}
Where $\Delta t$ is the interval of time that the motion of the agent takes.

This is for an individual trajectory. Summing over all trajectories, we get the total number of events, the flow, times the average time of one crossing for all agents. The sum of all times for all events is the number of events times the average time.  Then for identical agents, the denominator becomes:
\begin{equation}
   Q={\sum_{i=1}^{nm} I_{i,j}} \label{eq:10}=nmLt=\phi L\left\langle t \right\rangle
\end{equation}

Therefore:
\begin{equation}
   {\left\langle \alpha \right\rangle}= \frac{h \phi}{\phi L\left\langle t \right\rangle} \label{eq:11}
\end{equation}
and:
\begin{equation}
   {\left\langle \alpha \right\rangle}= \frac{h }{ L\left\langle t \right\rangle} \label{eq:12}
\end{equation}
The Lagrangian is just the kinetic energy, because the potential energy in this simulation is zero, and in the simulation, we are free to set the mass to two and the velocity is one patch per second. Therefore, we can have the kinetic energy to be equal to one. This equation is used for the calculations in this paper. 

Since Planck's constant is a fundamental unit of action, even though action can vary continuously, this equation represents how far is the organization of the system from this highly action-efficient state, when there will be only one Planck unit of action per event. The action itself can be even smaller than $h$ \citep{Feynman1948}. This provides a path to further continuous improvement in the levels of organization of systems below one quantum of action. 

\textbf{An example for one agent: }

To illustrate the simplest possible case, for clarity, we apply this model to the example of a closed system in two dimensions with only one agent. We define the boundaries of the fixed system to form a square. 

The endpoints here represent two nodes in a complex network. Thus the model is limited only to the path between the two nodes. The expansion of this model will be to include many nodes in the network and to average over all of them. Another extension is to include many elements, different kinds of elements, obstacles, friction, etc. 

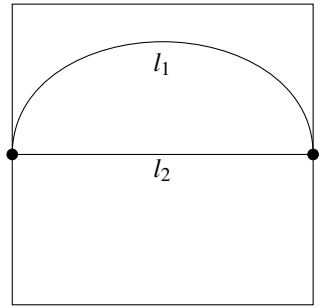
\begin{figure}[ht]
    \centering % This centers the figure
    \begin{subfigure}[b]{0.5\textwidth} % Adjust the width as needed
        \centering % This centers the subfigure
        \begin{tikzpicture}
        %a picture for my paper
        \draw (0,0) rectangle (4,-4); % Draw a rectangle
        \draw plot[mark=*, mark options={fill=black}] coordinates {(0,-2) (4,-2)};
        % Label for the shorter line (l2)
        \node at (2, -2.2) {$l_2$};
        % Draw a curved line from (0,0) to (4,0)
        \draw (0,-2) to[out=90,in=180] (2,-0.5) to[out=0,in=90] (4,-2);
        % Label for the longer curve (l1)
        \node at (2, -0.8) {$l_1$};
        \end{tikzpicture}
    \end{subfigure}
    \vspace{10pt} % Add some space between the figure and the caption
    \caption{Comparison between the geodesic and a longer path between two nodes in a network. }
\end{figure}

Figure A1 shows the boundary conditions for the system used in this example. In this figure, we present the boundaries of the system and define the initial and final points of the motion of an agent as two of the nodes in a complex network.

Comparison between two different states of organization of the system. This is a schematic representation of the two states of the system, and the shortest path of the agent in each case.  Here $l_1$ and $l_2$ are the lengths of the trajectory of the agent in each case. (a) a trajectory (geodesic) of an agent in a certain state of the system, given by the configuration of the internal constraints, $l_1$. (b) a different configuration allowing the geodesic trajectory of the element to decrease by 50\%, $l_2$  - the shortest possible path.

For this case, we set $n=1$, $m=1$, which is one crossing of one agent between two nodes in the network. An approximation for an isotropic medium (assumption \#8) allows us to express the time using the speed of the element when it is constant (Assumption \#3). In this case, then we can solve $v = \frac{l}{\Delta t}$ which is the definition of average velocity for the interval of time as $\Delta t = \frac{l}{v}$, where $l$ is the length of the trajectory of the element in each case between the endpoints.

The speed of the element $v$ is fixed to be another constant, so the action integral takes the form:
\begin{equation}
    I = L\Delta t = L \frac{l}{v}  \label{eq:14}
\end{equation}
When we substitute eq. \ref{eq:14} in the expression for organization, eq. X, we obtain:
\begin{equation}
    \alpha = \frac{\text{h}}{I} = \frac{hv}{L l}\label{eq:15}
\end{equation}

For the simulation in this paper, $\boldsymbol{l}$  is the distance that the ants travel between food and nest. Because h, v, and L are all constants, we can simplify this as we set 
\begin{equation}
    C =  \frac{hv}{L }\label{eq:16}
\end{equation}

And rewrite:

\begin{equation}
    \alpha =  \frac{hv}{L l} =\frac{C}{ l}\label{eq:17}
\end{equation}

We can set this constant to $C=1$, when necessary.

\subsection{Analysis of System States}
Now we turn to the two states of the system with different actions of the elements, as shown in \textbf{Fig. A1.} The organization of those two states is respectively: 
\begin{equation}
    \alpha_1 = \frac{C}{l_1} \text{ in state 1, and } \alpha_2 = \frac{C}{l_2} \text{ in state 2 of the system.}
\end{equation}
In \textbf{Fig. A1.} , the length of the trajectory in the second case (b) is less, $l_2 < l_1$, which indicates that state 2 has better organization. The difference between the organizations in the two states of the same system is generally expressed as:
\begin{equation}
    \alpha_2 - \alpha_1 = \frac{C}{l_2} - \frac{C}{l_1} = C \left( \frac{1}{l_2} - \frac{1}{l_1} \right) = C \left( \frac{l_1 - l_2}{l_1 l_2} \right) \label{eq:18}
\end{equation}

This can be rewritten as:
\begin{equation}
    \Delta \alpha = C \left( \frac{\Delta l}{\prod_{i=1}^{2} l_i} \right) \label{eq:19}
\end{equation}

Where $\Delta \alpha = \alpha_2 - \alpha_1$, $\Delta l = l_1 - l_2$, and $\prod_{i=1}^{2} l_i = l_1 l_2$.

This is for one agent in the system. If we describe the multi-agent system, then, we use average path-length.

\subsection{Average Action Efficiency (AAE)}

In the previous example, we can say that the shorter trajectory represents a more action-efficient state, in terms of how much total action is necessary for the event in the system, which here is for the agent to cross between the nodes. If we expand to many agents between the same two nodes, all with slightly different trajectories, we can define that the average of the action necessary for each agent to cross between the nodes is the average action efficiency. Average action efficiency is how efficiently a system utilizes energy and time to perform the events in the system. More organized systems are more action-efficient because they can perform the events in the system with fewer resources, in this example, energy and time.

We can start from the presumption that the average action efficiency in the most organized state is always greater than or equal to its value in any other configuration, arrangement, or structure of the system. By varying the configurations of the structure until the average action efficiency is maximized, we can identify the most organized state of the system. This state corresponds to the minimum average action per event in the system, adhering to the principle of least action. We refer to this as the ground or most stable state of the system, as it requires the least amount of action per event. All other states are less stable because they require more energy and time to perform the same functions. 

If we define average action efficiency as the ratio of useful output, here it is the crossing between the nodes, and, in other systems, it can be any other measure, to the total input or the energy and time expended, a system that achieves higher action efficiency is more organized. This is because it indicates a more coordinated, effective interaction among the system's components, minimizing wasted energy or resources for its functions.

During the process of self-organization, a system transitions from a less organized to a more organized state. If we monitor the action efficiency over time, an increase in efficiency could indicate that the system is becoming more organized, as its components interact in a more coordinated way and with fewer wasted resources. This way we can measure the level of organization and the rate of increase of action efficiency which is the level and the rate of self-organization, evolution, and development in a complex system. 

To use action efficiency as a quantitative measure, we need to define and calculate it precisely for the system in question. For example, in a biological system, efficiency might be measured in terms of energy conversion efficiency in cells. In an economic system, it can be the ratio of production of an item to the total time, energy, and other resources expended. In a social system, it could be the ratio of successful outcomes to the total efforts or resources expended. 

\textbf{The predictive power of the Principle of Least Action for Self-Organization:}\\ For the simplest example here of only two nodes, calculating theoretically the least action state as the straight line between the nodes we arrive at the same state as the final organized state in the simulation in this paper. This is the same result from minimizing action and from any experimental result. It results in the geodesic of the natural motion of objects. When there are obstacles to the motion of agents, the geodesic is a curve described by the metric tensor. To achieve this prediction for multiagent systems we minimize the average action between the endpoints. Therefore the most organized state in the current simulation is predicted theoretically from the principle of least action. Therefore, the Principle of Least Action provides a predictive power for calculating the most organized state of a system, and verifying it with simulations or experiments. In engineered or social systems, it can be used to predict the most organized state and then construct it. 

\subsection{Multi-agent}
Now we turn to the two states of the system with different average actions of the elements on Figure A1. The organization of those two states is respectively: 
\begin{equation}
    \left\langle \alpha \right\rangle_1 = \frac{C}{\left\langle l \right\rangle_1} \text{ in state 1, and } \left\langle \alpha \right\rangle_2 = \frac{C}{\left\langle l \right\rangle_2} \text{ in state 2 of the system.}
\end{equation}

The average length of the trajectories in the second case is less, $ \left\langle l \right\rangle_2 < \left\langle  l\right\rangle_1$, which indicates that state 2 has better organization. The difference between the organizations in the two states of the same system is generally expressed as:
\begin{equation}
   \left\langle \alpha \right\rangle_2 - \left\langle \alpha \right\rangle_1 = \frac{C}{\left\langle l \right\rangle_2} - \frac{C}{\left\langle l \right\rangle_1} = C \left( \frac{1}{\left\langle l \right\rangle_2} - \frac{1}{\left\langle l \right\rangle_1} \right) = C \left( \frac{\left\langle l \right\rangle_1 - \left\langle l \right\rangle_2}{\left\langle l \right\rangle_1 \left\langle l \right\rangle_2} \right) \label{eq:20}
\end{equation}

This can be rewritten as:
\begin{equation}
    \Delta \left\langle \alpha \right\rangle = C \left( \frac{\Delta \left\langle l \right\rangle}{\prod_{i=1}^{2} \left\langle l \right\rangle_i} \right) \label{eq:21}
\end{equation}

Where $\Delta \left\langle \alpha \right\rangle = \left\langle \alpha \right\rangle_2 - \left\langle \alpha \right\rangle_1$, $\Delta \left\langle l\right\rangle = \left\langle l \right\rangle_1 - \left\langle l \right\rangle_2$, and $\prod_{i=1}^{2} \left\langle l \right\rangle_i = \left\langle l \right\rangle_1 \left\langle l \right\rangle_2$.

This is when we use the average lengths of the trajectories and when the velocity is constant and the time and length are the same. In general, when the velocity varies we need to use time.

\subsection{Using time}
In this case, the two states of the system are with different average actions of the elements. The organization of those two states is respectively: 
\begin{equation}
    \left\langle \alpha \right\rangle_1 = \frac{C}{\left\langle t \right\rangle_1} \text{ in state 1, and } \left\langle \alpha \right\rangle_2 = \frac{C}{\left\langle t \right\rangle_2} \text{ in state 2 of the system.}
\end{equation}
In Fig. A1, the length of the trajectory in the second case (b) is less, the average time for the trajectories is $ \left\langle t \right\rangle_2 < \left\langle  t \right\rangle_1$, which indicates that state 2 has better organization. The difference between the organizations in the two states of the same system is generally expressed as:
\begin{equation}
   \left\langle \alpha \right\rangle_2 - \left\langle \alpha \right\rangle_1 = \frac{C}{\left\langle t \right\rangle_2} - \frac{C}{\left\langle t \right\rangle_1} = C \left( \frac{1}{\left\langle t \right\rangle_2} - \frac{1}{\left\langle t \right\rangle_1} \right) = C \left( \frac{\left\langle l \right\rangle_1 - \left\langle t \right\rangle_2}{\left\langle t \right\rangle_1 \left\langle t \right\rangle_2} \right) \label{eq:22}
\end{equation}

This can be rewritten as:
\begin{equation}
    \Delta \left\langle \alpha \right\rangle = C \left( \frac{\Delta \left\langle t \right\rangle}{\prod_{i=1}^{2} \left\langle t \right\rangle_i} \right) \label{eq:23}
\end{equation}

Where $\Delta \left\langle \alpha \right\rangle = \left\langle \alpha \right\rangle_2 - \left\langle \alpha \right\rangle_1$, $\Delta \left\langle t \right\rangle = \left\langle t \right\rangle_1 - \left\langle t \right\rangle_2$, and $\prod_{i=1}^{2} \left\langle t \right\rangle_i = \left\langle t \right\rangle_1 \left\langle t \right\rangle_2$.

Which, recovering C, is:
\begin{equation}
    \Delta \left\langle \alpha \right\rangle = \frac{hv}{L} \left( \frac{\Delta \left\langle t \right\rangle}{\prod_{i=1}^{2} \left\langle t \right\rangle_i} \right) \label{eq:24}
\end{equation}

\subsection{An Example}

For the simplest example of one agent and one crossing between two nodes if $l_1 = 2l_2$, or the first trajectory is twice as long as the second, this expression produces the result:
\begin{equation}
    \alpha_1 = \frac{C}{2l_2} = \frac{\alpha_2}{2} \text{ or } \alpha_2 = 2  \alpha_1,
\end{equation}
indicating that state 2 is twice as well organized as state 1. Alternatively, substituting in eq. \ref{eq:24} we have:
\begin{equation}
    \alpha_2 - \alpha_1 = C \left( \frac{2 - 1}{2} \right) = \frac{C}{2},
\end{equation}
or there is a 50\% difference between the two organizations, which is the same as saying that the second state is quantitatively twice as well organized as the first one. This example illustrates the purpose of the model for direct comparison between the amounts of organization in two different states of a system. When the changes in the average action efficiency are followed in time, we can measure the rates of self-organization. 

In our simulations, the higher the density and the lower the entropy of the agents, the shorter the paths and the time for crossing them, and the more the action efficiency.

\subsection{Unit-total (local-global) dualism}

In addition to the classical stationary action principle for fixed, non-growing, non-self-organizing systems:
\[\begin{aligned}\delta I = 0\\ 
\end{aligned}\]
we find a dynamical action principle: 
\[\begin{aligned}\delta I \neq 0\\ 
\end{aligned}\]
This principle exhibits a unit-total (local-global, min-max) dualism:

1. Average unit action for one edge decreases:

\[\begin{aligned}\delta \dfrac{\sum ^{n,m}_{i,j=1}I_{i,j}}{nm} <0\\
\end{aligned}\]

This is a principle for decreasing unit action for a complex system during self-organization, as it becomes more action-efficient until a limit is reached. 

2. Total action of the system increases:

\[\begin{aligned}\delta \sum ^{n,m}_{i,j=1}I_{i,j} >0\\
\end{aligned}\]

This is a principle for increasing total action for a complex system during self-organization, as the system grows until a limit is reached.

In our data, we see that average unit action, in terms of action efficiency decreases while total action increases Figure \ref{a-Q}. Both are related strictly with a power law relationship, predicted by the model of positive feedback between the characteristics of the system. 

Analogously, unit internal Boltzmann entropy for one path is decreasing while total internal Boltzmann entropy is increasing for a complex system during self-organization and growth Figure \ref{UnitEntVsFinEnt}. These two characteristics are also related strictly to a power law relationship, predicted by the model of positive feedback between the characteristics of the system. 

For the Gauss' principle of least constraint \citep{gauss1829neues} this will translate as the unit constraint (obstacles) for one edge decreases, the total constraints in the network of the whole complex system during self-organization increases as it grows and expands. 

For Hertz's principle of least curvature \citep{goldstein:mechanics}, this will translate as the unit curvature for one edge decreases, the total curvature in the network of the whole complex system during self-organization increases as it grows and expands and adds more nodes. 

Some examples of unit-total (local-global) dualism in other systems are: In economies of scale as the size of the system grows, the total production cost increases as the unit cost per one item decreases. In the same example, the total profits increase, but the unit profit per item decreases. Also, as the cost per one computation decreases, the cost for all computations grows. As the cost per one bit of data transmission decreases the cost for all transmissions increases as the system increases. In biology, as the unit time for one reaction in a metabolic autocatalytic cycle decreases in evolution, due to increased enzymatic activity, the total number of reactions in the cycle increases. In ecology, as one species becomes more efficient in finding food, its time and energy expenditure for foraging a unit of food decreases, the numbers of that species increase and the total amount of food that they collect increases. We can keep naming other unit-total (local-global) dualisms in systems of a very different nature, to test the universality of this principle. 
\section{Simulations Model}

In our simulation, the ants are interacting through pheromones. We can formulate an effective Lagrangian to describe their dynamics. The Lagrangian \( L \) depends on the kinetic energy \( T \) and the potential energy \( V \). We can start building it slowly by adding necessary terms to the Lagrangian. Given that ants are influenced by pheromone concentrations, the potential energy component should reflect this interaction.

 Components of the Lagrangian:
1. Kinetic Energy (\( T \)): In our simulation, the ants have a constant mass \( m \), and their kinetic energy is given by:
     \[
     T = \frac{1}{2} m v^2
     \]
     where \( v \) is the velocity of the ant.

2. Effective Potential Energy (\( V \)): The potential energy due to pheromone concentration \( C(\mathbf{r}, t) \) at position \( \mathbf{r} \) and time \( t \) can be modeled as:
     \[
     V_{\text{eff}} = -k C(\mathbf{r}, t)
     \]
     where \( k \) is a constant that scales the influence of the pheromone concentration.

Effective Lagrangian (\( L \)): The Lagrangian \( L \) is given by the difference between the kinetic and potential energies:
  \[
  L = T - V
  \]

For an ant moving in a pheromone field, the effective Lagrangian becomes:
\[
L = \frac{1}{2} m v^2 + k C(\mathbf{r}, t)
\]

Formulating the Equations of Motion:
 
Using the Lagrangian, we can derive the equations of motion via the Euler-Lagrange equation:
\[
\frac{d}{dt} \left( \frac{\partial L}{\partial \dot{x}_i} \right) - \frac{\partial L}{\partial x_i} = 0
\]
where \( x_i \) represents the spatial coordinates (e.g., \( x, y \)) and \( \dot{x}_i \) represents the corresponding velocities.

 Example Calculation for a Single Coordinate:
 
1. Kinetic Energy Term:
   \[
   \frac{\partial L}{\partial \dot{x}} = m \dot{x}
   \]
   \[
   \frac{d}{dt} \left( \frac{\partial L}{\partial \dot{x}} \right) = m \ddot{x}
   \]

2. Potential Energy Term:
   \[
   \frac{\partial L}{\partial x} = k \frac{\partial C}{\partial x}
   \]

The equation of motion for the \( x \)-coordinate is then:
\[
m \ddot{x} = k \frac{\partial C}{\partial x}
\]

Full Equations of Motion:
 
For both \( x \) and \( y \) coordinates, the equations of motion are:
\[
m \ddot{x} = k \frac{\partial C}{\partial x}
\]
\[
m \ddot{y} = k \frac{\partial C}{\partial y}
\]
The ants are moving following the gradient of the concentration.

 Testing for stationary Points of Action:
\begin{enumerate}
    \item  Minimum: If the second variation of the action is positive, the path corresponds to a minimum of the action.
    \item  Saddle Point: If the second variation of the action can be both positive and negative depending on the direction of the variation, the path corresponds to a saddle point.
    \item Maximum: If the second variation of the action is negative, the path corresponds to a maximum of the action.
\end{enumerate}

Determining the Nature of the Stationary Point:
 
To determine whether the action is a minimum, maximum, or saddle point, we examine the second variation of the action, \( \delta^2 I \). This involves considering the second derivative (or functional derivative in the case of continuous systems) of the action with respect to variations in the path.

Given the Lagrangian for ants interacting through pheromones

The action is:
\[
I = \int_{t_1}^{t_2} \left( \frac{1}{2} m \dot{\mathbf{r}}^2 + k C(\mathbf{r}, t) \right) dt
\]

First Variation:
 
The first variation \( \delta I \) leads to the Euler-Lagrange equations, which give the equations of motion:
\[
m \ddot{\mathbf{r}} = k \nabla C(\mathbf{r}, t)
\]

Second Variation:
 
The second variation \( \delta^2 I \) determines the nature of the stationary point. In general, for a Lagrangian \( L = T - V \):
\[
\delta^2 I = \int_{t_1}^{t_2} \left( \delta^2 T - \delta^2 V \right) dt
\]

 Analyzing the Effective Lagrangian:
 
1. Kinetic Energy Term \( T = \frac{1}{2} m \dot{\mathbf{r}}^2 \): The second variation of the kinetic energy is typically positive, as it involves terms like \( m (\delta \dot{\mathbf{r}})^2 \).

2. Potential Energy Term \( V_{\text{eff}} = -k C(\mathbf{r}, t) \): The second variation of the effective potential energy depends on the nature of \( C(\mathbf{r}, t) \). If \( C \) is a smooth, well-behaved function, the second variation can be analyzed by examining \( \nabla^2 C \).

 Nature of the Stationary Point:
 
\begin{enumerate}
    \item Kinetic Energy Contribution: Positive definite, contributing to a positive second variation.
    \item Effective Potential Energy Contribution: Depends on the curvature of \( C(\mathbf{r}, t) \). If \( C(\mathbf{r}, t) \) has regions where its second derivative is positive, the effective potential energy contributes positively, and vice versa.
\end{enumerate}

Therefore, given the typical form of the Lagrangian and assuming \( C(\mathbf{r}, t) \) is well-behaved (smooth and not overly irregular), the action \( S \) is most likely a saddle point. This is because:

\begin{enumerate}
    \item The kinetic energy term tends to make the action a minimum.
    \item The potential energy term, depending on the pheromone concentration field, can contribute both positively and negatively.
\end{enumerate}

Thus, variations in the path can lead to directions where the action decreases (due to the kinetic energy term) and directions where it increases (due to the potential energy term), characteristic of a saddle point.

Incorporating factors such as the wiggle angle of ants and the evaporation of pheromones introduces additional dynamics to the system, which can affect whether the action remains stationary, a saddle point, a minimum, or a maximum. Here's how these changes influence the nature of the action:

Effects of Wiggle Angle and Pheromone Evaporation on the Action

1. Wiggle Angle: Impact: The wiggle angle introduces stochastic variability into the ants' paths. This randomness can lead to fluctuations in the paths that ants take, affecting the stability and stationarity of the action.
   Mathematical Consideration: The additional term representing the wiggle angle’s variance in the Lagrangian adds a stochastic component:
     \[
     L = \frac{1}{2} m v^2 + k C(\mathbf{r}, t) + \frac{1}{2} \sigma^2 (\theta)
     \]
    Consequence: The action is less likely to be strictly stationary due to the inherent variability introduced by the wiggle angle. This can lead to more dynamic behavior in the system.

2. Pheromone Evaporation:
   Impact: Pheromone evaporation reduces the concentration of pheromones over time, making previously attractive paths less so as time progresses.
   Mathematical Consideration: Including the evaporation term in the Lagrangian:
     \[
     L = \frac{1}{2} m v^2 + k C(\mathbf{r}, t) e^{-\lambda t}
     \]
Consequence: The time-dependent decay of pheromones means that the action integral changes dynamically. Paths that were optimal at one point may no longer be optimal later, leading to continuous adaptation.

\subsection{Considering the Nature of the Action}

Given these modifications, the nature of the action can be characterized as follows:

1. Stationary Action:
   - Before Changes: In a simpler model without wiggle angles and evaporation, the action might be stationary at certain paths.
   - After Changes: With wiggle angle variability and pheromone evaporation, the action is less likely to be stationary. Instead, the system continuously adapts, and the action varies over time.

2. Saddle Point, Minimum, or Maximum:
   - Saddle Point: The action is likely to be at a saddle point due to the dynamic balancing of factors. The system may have directions in which the action decreases (due to pheromone decay) and directions in which it increases (due to path variability).
   - Minimum: If the system stabilizes around a certain path that balances the stochastic wiggle and the decaying pheromones effectively, the action might approach a local minimum. However, this is less likely in a highly dynamic system.
   - Maximum: It is unusual for the action in such optimization problems to represent a maximum because that would imply an unstable and inefficient path being preferred, which is contrary to observed behavior.

Practical Implications

1. Continuous Adaptation:
   - The system will require continuous adaptation to maintain optimal paths. Ants need to frequently update their path choices based on the real-time state of the pheromone landscape.

2. Complex Optimization:
   - Optimization algorithms must account for the random variations in movement, the rules for deposition and diffusion and the temporal decay of pheromones. This means more sophisticated models and algorithms are necessary to predict and find optimal paths.

Therefore, incorporating the wiggle angle and pheromone evaporation into the model makes the action more dynamic and less likely to be strictly stationary. Instead, the action is more likely to exhibit behavior characteristic of a saddle point, with continuous adaptation required to navigate the dynamic environment. This complexity necessitates advanced modeling and optimization techniques to accurately capture and predict the behavior of the system.

\subsection{Dynamic Action}

For dynamical non-stationary action principles, we can extend the classical action principle to include time-dependent elements. The Lagrangian is changing during the motion of an agent between the nodes as the terms in it are changing. 

	1.	Time-Dependent Lagrangian that explicitly depends on time or other dynamic variables:
\[
L = L(q, \dot{q}, t, \lambda(t))
\]
where ( q ) represents the generalized coordinates, ( $\dot{q}$) their time derivatives, ( t ) time, and ($\lambda(t)$ ) a set of dynamically evolving parameters.
	2.	Dynamic Optimization - the system continuously adapts its trajectory  q(t)  to minimize or optimize the action that evolves over time:
\[
I = \int_{t_1}^{t_2} L(q, \dot{q}, t, \lambda(t)) \, dt
\]

	The parameters $ \lambda(t) $ are updated based on feedback from the system’s performance. The goal is to find the path \( q(t) \) that makes the action stationary. However, since \( \lambda(t) \) is time-dependent, the optimization becomes dynamic.

Euler-Lagrange Equation

To find the stationary path, we derive the Euler-Lagrange equation from the time-dependent Lagrangian. For a Lagrangian \( L(q, \dot{q}, t, \lambda(t)) \), the Euler-Lagrange equation is:

\[ \frac{d}{dt} \left( \frac{\partial L}{\partial \dot{q}} \right) - \frac{\partial L}{\partial q} = 0 \]

However, due to the dynamic nature of \( \lambda(t) \), additional terms may need to be considered.

Updating Parameters \( \lambda(t) \)

The parameters \( \lambda(t) \) evolve based on feedback from the system’s performance. This feedback mechanism can be modeled by incorporating a differential equation for \( \lambda(t) \):

\[ \frac{d\lambda(t)}{dt} = f(\lambda(t), q(t), \dot{q}(t), t) \]

Here, \( f \) represents a function that updates \( \lambda(t) \) based on the current state \( q(t) \), the velocity \( \dot{q}(t) \), and possibly the time \( t \). The specific form of \( f \) depends on the nature of the feedback and the system being modeled.

Practical Implementation

In our example of ants with a wiggle angle and pheromone evaporation. The effective Lagrangian will look like this:

\[ L = \frac{1}{2} m v^2 + k C(\mathbf{r}, t) e^{-\lambda(t) t} + \frac{1}{2} \sigma^2 (\theta) \]

- \( \frac{1}{2} m v^2 \): Kinetic energy of the ant.
- \( k C(\mathbf{r}, t) e^{-\lambda(t) t} \): Pheromone concentration at position \( \mathbf{r} \) and time \( t \), decaying exponentially with rate \( \lambda(t) \).
- \( \frac{1}{2} \sigma^2 (\theta) \): Variance in the wiggle angle \( \theta \), representing the stochastic movement of the ant.

The action \( I \) would be:

\[ I = \int_{t_1}^{t_2} \left( \frac{1}{2} m v^2 + k C(\mathbf{r}, t) e^{-\lambda(t) t} + \frac{1}{2} \sigma^2 (\theta) \right) dt \]

 Dynamical System Adaptation

The system adapts by updating \( \lambda(t) \) based on the current state of pheromones and the ants’ paths. For example, \( \lambda(t) \) could represent the pheromone evaporation rate, or on the density of ants reinforcing the pheromone trail.

Solving the Equations

1. Numerical Methods: Usually, these systems are too complex for analytical solutions, so numerical methods (e.g., finite difference methods, Runge-Kutta methods) are used to solve the differential equations governing \( q(t) \) and \( \lambda(t) \).
2. Optimization Algorithms: Algorithms like gradient descent, genetic algorithms, or simulated annealing can be used to find optimal paths and parameter updates.

By extending the classical action principle to include time-dependent and evolving elements, we can model and solve more complex, dynamic systems. This framework is particularly useful in real-world scenarios where conditions change over time, and systems must adapt continuously to maintain optimal performance. This approach is applicable in physical, chemical, and biological systems, and in fields such as robotics, economics, and ecological modeling, providing a powerful tool for understanding and optimizing dynamic, non-stationary systems.

The lagrangian changes at each time step of the simulation, therefore we cannot talk about static action, but a dynamic action. This is dynamic optimization and reinforcement learning. 

The average action is quasi-stationary, as is fluctuates around a fixed value, but, internally, each trajectory which it is composed of is fluctuating stochastically given the dynamic lagrangian of each ant. It still fluctuates around the shortest theoretical path, so the average action is minimized far from the stationary path, even though close to the minimum it can be stuck in a neighboring stationary action path temporarily. In all these situations, as described above, the average action efficiency is our measure for organization. 

\subsection{Specific details in our simulation}
For our simulation the details of the concentration changes at each patch $C(r,t)$ at each update are the sum of three contributions and can be included as:

1. $C_{i,j}(t)$ is the preexisting amount of pheromone at each patch at time t.

2. Pheromone Diffusion: The changes of the pheromone at each patch at time $t$, are described by the rules of the simulation: 70\% of the pheromone is split between all 8 neighboring patches on each tick, regardless of how much pheromone is in that patch, which means that 30\% of the original amount is left in the central patch. On the next tick 70\% of those remaining 30\% will diffuse again. At the same time, using the same rule, pheromone is distributed from all 8 neighboring ants to the central one. Note: this rule for diffusion does not follow the diffusion equations in physics, where there is always flow from high concentration to low. 

\[ C_{i,j}(t+1)=0.3C_{i,j}(t)+\frac{0.7}{8}\left( {\sum_{k,l=-1}^{1}{C_{i+k, j+l}(t)}}\right) \]where $|k|+|l| \neq 0$

The first term in the equation shows how much of the concentration of the pheromone from the previous time step is left in the next, and the second term shows the incoming pheromone from all neighboring patches, as 70\% 1/8 of each concentration is distributed to the central one. 

3. The amount of pheromone an ant deposits after n steps can be expressed as: 
\[
P(n)=\frac{1}{10}P_{0}(0.9)^n
\]
Where $P(0)=30$

The stochastic Term (\(\sigma^2(\theta)\)) for the parameters in this simulation, is the variance of a uniform distribution\citep{libretexts_uniform_distribution}:
   \[
   \sigma^2 (\theta) = \frac{50^2}{12}
   \]

\subsection{Gradient based approach}

We can use either the concentration's value or the concentration gradient in the potential energy term. Using the gradient is a more exact approach but even more computationally intensive. 

In further extension of the model, we can incorporate a gradient-based potential energy term. In this case, the Lagrangian becomes:

\[
L = \frac{1}{2} m v^2 + k \left| \nabla C(\mathbf{r}, t) \right| + \frac{1}{2} \left( \frac{50^2}{12} \right)
\]

\subsection{Summary}

1. We obtained the Lagrangian with the exact parameters for the specific current simulation that produced the data. Up to our current knowledge we don't know of other studies which have published the lagrangian approach to agent based simulations of ants.

2. The lagrangian is impossible to solve analytically, to our current knowledge, due to the stochastic term 

The equation for the concentration of pheromones is at a given patch, but the equation for the amount deposited by the ants depends on how many steps, n, they have taken since they visited the food or nest. Each ant has a different path, so n in the equation will be different for each ant and it will be depositing a different amount of pheromones. This in general cannot be done analytically, it is dependent on the stochastic paths of each ant, therefore the only way to solve it numerically through the simulation. Also, in the equation, the concentration is for each patch i,j. We solve it numerically through the simulation.

3.  The average path length obtained from the simulation serves as a numerical solution to the Lagrangian because it results from the model that incorporates all the dynamics described by the Lagrangian. This path length reflects the optimization and behaviors modeled by the Lagrangian terms, including kinetic energy, potential energy influenced by pheromone concentrations, and stochastic movement. The simulation is using $[\frac{1}{average-path-length}]$  as the average action efficiency. This takes into account all of the effects on the Lagrangian. The average path length is the final result. 

4. The average action could be stationary close to the theoretically shortest path, i.e. near the minimum of the average action, but further away from it it is always minimized, experimentally and from theoretical considerations. In the simulation, it is measured that longer paths always decay to shorter paths. There can only be some deviations very close to the shortest path due to memory effects and stochastic reasons which will decay with longer annealing and changing parameters such as exploration by increasing the wiggle angle, changing the pheromone deposition, diffusion and evaporation rates, changing the speed and mass of the ants, and other factors. When the average action efficiency is growing it means that the average unit action is decreasing. When the action is stationary, as it is at the end of the simulations, as seen in the time graphs, the average action efficiency is also stationary - it does not grow anymore, in the time and for the parameters of the current simulation. A similar process is happening in real complex systems such as organisms, ecosystems, cities, and economies. Due to the stochastic variations, we can consider only average quantities.

\section{Mechanism}

\subsection{Exponential Growth and Size-Complexity Rule}

Average action efficiency is the proposed measure for level of organization and complexity. To test it we turn to observational data. The size-complexity rule states that the complexity increases as a power law of the size of a complex system \citep{Bonner2004size}. This rule has been observed in systems of a very different nature, without explanation or proposed origin. In the next section on the model of the mechanism of self-organization, we derive those exponential and power law dependencies. In this paper, we show how our data aligns with the size-complexity rule.

\subsection{A model for the mechanism of self-organization}

We apply the model first presented in our paper from 2015 \citep{georgiev2015mechanism} and used in the following papers \citep{georgiev2017exponential, Georgiev2021Stars} to the ABM simulation here, and specify all of the quantities in this model. Then, we show the exponential and power law solutions for this specific system. 

Below is a visual representation of the positive feedback interactions between the characteristics of a complex system, which in our 2015 paper \citep{georgiev2015mechanism} have been proposed as the mechanism of self-organization, progressive development, and evolution, applied to the current simulation.  Here $i$ is the information in the system, calculated by the total amount of ant pheromones, $\left\langle t\right\rangle$ is the average time for all of the ants in the simulation crossing between the two nodes, $N$ is the total number of ants, $Q$ is the total action of all ants in the system, $\Delta S$ is the internal entropy difference between the initial and final state of the system in the process of self-organization of finding the shortest path, $\left\langle \alpha \right\rangle$ is the average action efficiency, $\phi$ is the number of events in the system, which in the simulation is the number of paths or crossings between the two nodes, $\Delta \rho$ is the order parameter, which is the difference in the density of agents between the final and initial state of the simulation. The links connecting all those quantities represent positive feedback connections between them. 

\begin{figure}[H]
    \centering
    \begin{tikzpicture}[auto,
                        thick,main node/.style={circle,draw,font=\sffamily\Large\bfseries}]
    
      \node[main node] (I) at (6,1) {i};
      \node[main node] (A) at (0,1) {$\left\langle t\right\rangle$};
      \node[main node] (N) at (0,-1) {N};
      \node[main node] (Q) at (2,-2) {Q};
      \node[main node] (DS) at (2,2) {$\Delta S$};
      \node[main node] (Alpha) at (4,-2) {$\left\langle \alpha \right\rangle$};
      \node[main node] (Phi) at (6,-1) {$\phi$};
      \node[main node] (Rho) at (4,2) {$\Delta \rho$};
    
      \foreach \source/ \dest in {I/A, I/N, I/Q, I/DS, I/Alpha, I/Phi, I/Rho,
                                  A/N, A/Q, A/DS, A/Alpha, A/Phi, A/Rho,
                                  N/Q, N/DS, N/Alpha, N/Phi, N/Rho,
                                  Q/DS, Q/Alpha, Q/Phi, Q/Rho,
                                  DS/Alpha, DS/Phi, DS/Rho,
                                  Alpha/Phi, Alpha/Rho,
                                  Phi/Rho}
        \path (\source) edge (\dest);
        %\vspace{10pt} % Add some space between the figure and the caption
    \end{tikzpicture}\\
    \label{fig1}
    \caption{Positive feedback model between the eight quantities in our simulation.}
\end{figure}
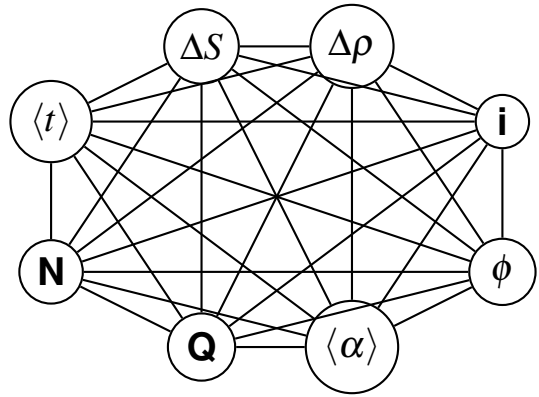

The positive feedback loops in Figure 1 are modeled with a set of ordinary differential equations. The solutions of this model are exponential for each characteristic and have a power law dependence between each two. The detailed solutions of this model are shown next.

\section{Mechanism}
This is the mathematical representation and solutions of the mechanism represented as a positive feedback loop between the eight characteristics of the system. 

In general, in a linear system with eight quantities, the shortest way to represent the interactions is by linear differential equations, using a matrix to describe the interactions between different quantities. We are writing this system generally in order to specify and discuss different aspects of it. Let's define our system as follows: 

\[
\frac{d}{dt}\begin{bmatrix}
x_1 \\
x_2 \\
x_3 \\
x_4 \\
x_5 \\
x_6 \\
x_7 \\
x_8
\end{bmatrix}
=
\begin{bmatrix}
a_{11} & a_{12} & a_{13} & a_{14} & a_{15} & a_{16} & a_{17} & a_{18} \\
a_{21} & a_{22} & a_{23} & a_{24} & a_{25} & a_{26} & a_{27} & a_{28} \\
a_{31} & a_{32} & a_{33} & a_{34} & a_{35} & a_{36} & a_{37} & a_{38} \\
a_{41} & a_{42} & a_{43} & a_{44} & a_{45} & a_{46} & a_{47} & a_{48} \\
a_{51} & a_{52} & a_{53} & a_{54} & a_{55} & a_{56} & a_{57} & a_{58} \\
a_{61} & a_{62} & a_{63} & a_{64} & a_{65} & a_{66} & a_{67} & a_{68} \\
a_{71} & a_{72} & a_{73} & a_{74} & a_{75} & a_{76} & a_{77} & a_{78} \\
a_{81} & a_{82} & a_{83} & a_{84} & a_{85} & a_{86} & a_{87} & a_{88} \\
\end{bmatrix}
\begin{bmatrix}
x_1 \\
x_2 \\
x_3 \\
x_4 \\
x_5 \\
x_6 \\
x_7 \\
x_8
\end{bmatrix}
\]

Here, \(\frac{d}{dt}\) denotes the derivative with respect to time, \(x_1, x_2, \ldots, x_8\) are the quantities of interest, and \(a_{ij}\) are constants that represent the interaction strengths between the quantities. The solutions for this system are exponential growth for each of the quantities, and power-law relationships can be derived from their exponential growth. Let's consider eight quantities \(x_1(t), x_2(t), ..., x_8(t)\) each growing exponentially:
\\[10pt] 
1. \(x_1(t) = x_{10} e^{a_1 t}\)\\
2. \(x_2(t) = x_{20} e^{a_2 t}\)\\
3. \(x_3(t) = x_{30} e^{a_3 t}\)\\
4. \(x_4(t) = x_{40} e^{a_4 t}\)\\
5. \(x_5(t) = x_{50} e^{a_5 t}\)\\
6. \(x_6(t) = x_{60} e^{a_6 t}\)\\
7. \(x_7(t) = x_{70} e^{a_7 t}\)\\
8. \(x_8(t) = x_{80} e^{a_8 t}\)\\

Each \(x_{i0}\) is the initial value, and each \(a_i\) is the growth rate for quantity \(x_i(t)\). To find a power law relationship between any two quantities, say \(x_i(t)\) and \(x_j(t)\):

1. Solve for \(t\) in terms of \(x_i(t)\) and \(x_j(t)\):

   \( t = \frac{1}{a_i} \ln\left(\frac{x_i(t)}{x_{i0}}\right) \)

   \( t = \frac{1}{a_j} \ln\left(\frac{x_j(t)}{x_{j0}}\right) \)

2. Set these two expressions equal to each other and solve for one variable in terms of the other:

   \(\frac{1}{a_i} \ln\left(\frac{x_i(t)}{x_{i0}}\right) = \frac{1}{a_j} \ln\left(\frac{x_j(t)}{x_{j0}}\right)\)

   \(\ln\left(\frac{x_i(t)}{x_{i0}}\right) = \frac{a_i}{a_j} \ln\left(\frac{x_j(t)}{x_{j0}}\right)\)

   \(\frac{x_i(t)}{x_{i0}} = \left(\frac{x_j(t)}{x_{j0}}\right)^{\frac{a_i}{a_j}}\)

   \(x_i(t) = x_{i0} \left(\frac{x_j(t)}{x_{j0}}\right)^{\frac{a_i}{a_j}} x_{j0}^{\frac{a_i}{a_j}}\)\\

This gives us a relationship between any two of the quantities \(x_i(t)\) and \(x_j(t)\). Now, replacing the variables, system of linear differential equations represented in matrix form becomes:

\[
\frac{d}{dt}\begin{bmatrix}
i \\
\left\langle t \right\rangle\\
N \\
Q \\
\Delta S \\
\left\langle \alpha \right\rangle \\
\varphi \\
\rho
\end{bmatrix}
=
A
\begin{bmatrix}
i \\
\left\langle t \right\rangle\\
N \\
Q \\
\Delta S \\
\left\langle \alpha \right\rangle \\
\varphi \\
\rho
\end{bmatrix}
\]

Here, \(A\) is the matrix of coefficients that define the interactions between the different quantities. For example, a list of some of the power-law-like relationships involving \(\alpha\) and \(Q\) with respect to the other variables based on their exponential growth relationships. Here we show only the relationships for average action efficiency and for total action:
\\
Relationships Involving \(\left\langle \alpha(t) \right\rangle \):\\
\\
1. \(\left\langle \alpha(t) \right\rangle = \alpha_0 \left(\frac{i(t)}{i_0}\right)^{\frac{a_6}{a_1}} i_0^{\frac{a_6}{a_1}}\)\\
2. \(\left\langle \alpha(t) \right\rangle = \alpha_0 \left(\frac{\left\langle t(t) \right\rangle}{\left\langle t \right\rangle_0}\right)^{\frac{a_6}{a_2}} \left\langle t \right\rangle _0^{\frac{a_6}{a_2}}\)\\
3. \(\left\langle \alpha(t) \right\rangle = \alpha_0 \left(\frac{N(t)}{N_0}\right)^{\frac{a_6}{a_3}} N_0^{\frac{a_6}{a_3}}\)\\
4. \(\left\langle \alpha(t) \right\rangle = \alpha_0 \left(\frac{Q(t)}{Q_0}\right)^{\frac{a_6}{a_4}} Q_0^{\frac{a_6}{a_4}}\) \\
5. \(\left\langle \alpha(t) \right\rangle = \alpha_0 \left(\frac{\Delta S(t)}{\Delta S_0}\right)^{\frac{a_6}{a_5}} \Delta S_0^{\frac{a_6}{a_5}}\)\\
6. \(\left\langle \alpha(t) \right\rangle = \alpha_0 \left(\frac{\varphi(t)}{\varphi_0}\right)^{\frac{a_6}{a_7}} \varphi_0^{\frac{a_6}{a_7}}\)\\
7. \(\left\langle \alpha(t) \right\rangle = \alpha_0 \left(\frac{\rho(t)}{\rho_0}\right)^{\frac{a_6}{a_8}} \rho_0^{\frac{a_6}{a_8}}\)\\
\\
 Relationships Involving \(Q(t)\):\\
 \\
1. \(Q(t) = Q_0 \left(\frac{i(t)}{i_0}\right)^{\frac{a_4}{a_1}} i_0^{\frac{a_4}{a_1}}\)\\
2. \(Q(t) = Q_0 \left(\frac{\left\langle t(t) \right\rangle}{\left\langle t \right\rangle_0}\right)^{\frac{a_4}{a_2}} \left\langle t \right\rangle_0^{\frac{a_4}{a_2}}\)\\
3. \(Q(t) = Q_0 \left(\frac{N(t)}{N_0}\right)^{\frac{a_4}{a_3}} N_0^{\frac{a_4}{a_3}}\)\\
4. \(Q(t) = Q_0 \left(\frac{\Delta S(t)}{\Delta S_0}\right)^{\frac{a_4}{a_5}} \Delta S_0^{\frac{a_4}{a_5}}\)\\
5. \(Q(t) = Q_0 \left(\frac{\left\langle \alpha(t) \right\rangle}{\alpha_0}\right)^{\frac{a_4}{a_6}} \alpha_0^{\frac{a_4}{a_6}}\) \\
6. \(Q(t) = Q_0 \left(\frac{\varphi(t)}{\varphi_0}\right)^{\frac{a_4}{a_7}} \varphi_0^{\frac{a_4}{a_7}}\)\\
7. \(Q(t) = Q_0 \left(\frac{\rho(t)}{\rho_0}\right)^{\frac{a_4}{a_8}} \rho_0^{\frac{a_4}{a_8}}\)\\

These equations describe how \(\alpha\) and \(Q\) scale with respect to each other and the other variables in the system, assuming all variables grow exponentially over time.

In our data, we see small deviations from the strict power law fits. A power-law can include a deviation term, which may show uncertainty in the values (measurement or sampling errors) or deviation from the power-law function (for example, for stochastic reasons):
\[ y = k x^n + \epsilon \]

where:
\begin{itemize}
    \item \( y \) and \( x \) are the variables.
    \item \( k \) is a constant.
    \item \( n \) is the exponent.
    \item \( \epsilon \) is a term that accounts for deviations. 
\end{itemize}

\section{Simulation Methods}

\subsection{Agent-Based Simulations approach}

Our study examines the properties of self-organization by simulating an ant colony navigating between a food source and its nest. The ants start with a random distribution, and then their trajectories become more correlated as they form a path. The ants pick up pheromone from the food and nest and lay it on each patch when they move. The food and the nest function as two nodes that attract the ants which follow the steepest gradient of the opposite pheromone. The pheromone is equivalent to information, as forming a path requires enough of it to ensure that the ants are able to follow the quickest path to their destinations rather than moving randomly. The ants can represent any agent in complex systems, from atoms and molecules to organisms, people, cars, dollars, or bits of information. Utilizing NetLogo for agent-based modeling and Python for data visualization and analysis, we measure self-organization via the calculated entropy decrease in the system, density order parameter, and average path length, which are contingent on the ants' distribution and possible microstates in the simulated environment. We further explore the effects of having different numbers of ants in the system simulating the growth of the system. In this study we look only at the final values of the characteristics at the end of the simulation when the self-organization is complete, or the difference between their values in the final and initial state. Then, we can demonstrate the relationship between each two characteristics as the population increases. Our model predicts that, according to one of the mechanisms of self-organization, evolution, and development of complex systems, all of the characteristics of a complex system reinforce each other, grow exponentially in time, and are proportional to each other by a power-law relationship \cite{georgiev2015mechanism}. The principle of least action is proposed as a driver for this process. The tendency to minimize action for crossing between the nodes in a complex network is a reason for self-organization and the average action efficiency is a measure of how organized the system is.

This simulation can utilize variables that affect the world, making it easier or harder to form the path. In the collected data, only the number of ants was changed. Increasing the number of ants makes it more probable to find the path, as there is not only a higher chance of them reaching the food and nest and adding information to the world, but also a steeper gradient of pheromone. This both increases the rate of path formation and decreases the length of the path. The ants follow the direction of the steepest gradient around them, but, their speed does not depend on how steep is the gradient. 

\subsection{Program Summary}
The simulation is run using an agent-based software called NetLogo. In the simulation, a population of ants forms a path between two endpoints, called the food and nest. The world is a 41x41 patch grid with 5x5 food and nest centered vertically on opposite sides and aligned with the edge. To help with path formation, there is a pheromone laid by the ants on a grid whenever the food or nest is reached. This pheromone exhibits the behavior of evaporating and diffusing across the world. The settings for ants and pheromones can be configured to make path formation easier or harder.

Each tick of the simulation functions as time, which represents a second in our simulation, according to the following rules. First, the ants check if there is any pheromone in its neighboring patches that are in a view cone with an angle of 135 degrees, oriented towards its direction of movement. From the position which the ant is in the current patch, it faces the center of the neighboring patch with the largest value of the pheromone in its viewing angle. It is important to note that the minimum amount of pheromone an ant can detect is $1/M$, where $M$ is the maximum amount of pheromone an ant can have, which in this simulation is 30. If there is not enough pheromone found in view, then the ant checks all neighboring patches with the same limitation for minimum pheromone. If any pheromone is found, if faces towards the patch with the highest amount. The ant then wiggles a random amount within an integer interval of -25 to 25 degrees, regardless of whether it faced any pheromone, and moves forward at a constant speed of 1 patch per tick. If the ant collides with the edge of the world, it turns 180 degrees and takes another step. In this simulation, the ants do not collide with obstacles or with each other. After it finishes moving, the ant checks if there is any food or nest in its current patch. A collision with the nest if the ant has food, or with the food source if it does not, will switch its status of having food, set its pheromone to 30, and update the path-length data. After the collision checks, it drops 1/10 of its current amount of pheromone at the patch. When all the ants have been updated, the patch pheromone is updated. There is a diffusion rate of 0.7, which means that 70\% of the pheromone at each patch is distributed equally to neighboring patches. There is also an evaporation rate of 0.06, which means that the pheromone at each patch is decreased by 6 percent. There are more behaviors than these available in the simulation.

\subsection{Analysis Summary}
The program stores information about the status of the simulation on each tick of the simulation. Upon completion of one simulation, the data is exported directly from the program for analysis by Python. Some of the data, such as Average Action Efficiency, is not directly exported from the program but must be generated by Python from other datasets. The data are fit with a power law function in python. To generate the graphs, the matplotlib Python library is used. The data seen in the graphs is the average of 20 simulations and has a moving average with a window of 50. Furthermore, any graph that requires the final value in the dataset obtains the value by averaging the last 200 points of the dataset without the moving average.

\subsection{Average Path Length}
The average path length, <l>, estimates the average length of the paths between food and nest created by the ants. On each tick, the path-length variable for each ant is increased by the amount by which it moved, which is 1 patch per tick for this simulation. When an ant reaches an endpoint, the path-length variable is stored in a list and reset to zero. This list is for all of the paths completed on that tick, and at the end of the tick, the list is averaged and added to the average path length dataset. If no paths were created, 0 is added to the average path length to serve as a placeholder; this can easily be removed in the analysis step because it is known that the path length cannot reach a length of 0. It is also important to note that, due to the method used to calculate this dataset, there will be a clear peak if a stable path is formed. This is because the path length of all the ants must begin at zero, the dataset is not representative before the peak, because the shorter paths from the start of the simulation are averaged. The peak itself shows a shifting trend with self-organization when parameters are changed. The average path length data in this simulation is identical to the average path time and can be used interchangeably whenever time is needed instead of distance. If the speed was varying, then the distance and the time would be different. 

\subsection{Flow Rate}
The flow rate, $\phi$, is the number of paths completed at each tick, or crossing between the nodes in this simple network, which is the number of events in the system, defined that way. It is the measure of how many ants reached the endpoints on each tick. This can simply be measured by counting how many ants reach the food or nest, and adding this value to the dataset. In this measure, there are a lot of fluctuations, so a moving average is necessary to make the graph readable.

\subsection{Final Pheromone}
The final pheromone is the total amount of pheromone at the end of the simulation, which is information for the ants. The amount of pheromone can be calculated on each tick by summing all the pheromone values from the nest and the food for each patch which vary during the simulation. By final pheromone, we mean the average of the pheromone for the last 200 ticks at the end of the simulation, 

\subsection{Total Action}
Action is calculated as the energy used times the time for each trajectory. Since energy is constant during the motion, it can be set to 1, so the individual action becomes equal to the time for one edge crossing, which is equal to the length of the trajectory. To get the total action for all agents, it is multiplied times the number of all events, or all crossings. The calculation for total action is based on flow rate and average path length. It is calculated after the simulation in Python using the equation $Q=\phi*<l>$. 

\subsection{Average Action Efficiency}
The definition for average action efficiency <$\alpha$> is the average amount of action per one event in the system or for one edge crossing. This is calculated by dividing the number of events by the total action in the system. The calculation for action efficiency is based on the data for average path length. It is calculated by the equation <$\alpha$>=1/<l>. This is based on the formula <$\alpha$>=$\phi$/Q. Note that the calculation for average action efficiency is first performed on the individual datasets, then the modified datasets are averaged, rather than averaging the datasets, then applying the equation.  

\subsection{Density}
The density of the ants can be used as an order parameter. To calculate the density of the ants, the simulation must get the average of how many ants are in each patch. To achieve this, the simulation approximates a box around the ants in the system to represent the area that they occupy. First, the center of the box is calculated by the equation $C_{x,y}=\langle p_{x,y}\rangle$for all ants, where $p_{x,y}$ is the position of each at each tick. Then, the length and width of the box are calculated by $S_{x,y}=4*\sqrt{\langle(p_{x,y}-C_{x,y})^{2}\rangle}$. Finally, the area can be calculated with the formula $A=S_{x}*S_{y}$. By using this  method of averaging the dimensions of the box instead of simply taking the furthest ant, it is ensured that a group of ants has priority over a few outliers. In Python, after the simulation is finished, the density of ants per patch can be calculated for each population, $N$, by $\rho=N/A$. It is the total number of ants divided by the area of the box in which the ants are concentrated. At the beginning of the simulation, the box takes the whole world, and as the ants form the path, it gradually decreases in size, corresponding to an increased density. 

\subsection{Entropy}
The system starts with maximum internal entropy, which decreases as paths are formed over time. The calculation for entropy is similar to the calculation of density. First, using the same method as described with density, a box is calculated around the ants within the system. 

We consider the agents in our simulation to be distinguishable because we have two different types of ants and each ant in the simulation is labeled and identifiable. The Boltzmann entropy is $S=k_B*ln(W)$. 

Where the number of states $W$ is the area $A$ that they occupy to the power of the number of ants $N$. 

\[ W  = {A^N} \]

Plugging this into the Boltzmann formula, we get:

\[ S = k_B \ln\left({A^N}\right) \]
Setting
\[k_B=1\]

We obtained the expression which we used in our calculations: 
\[ S =  N  \ln\left({A}\right)  \]
The box is the average size of the area $A$  in which the ants move. As the box decreases, the number of possible microstates in which they can be decreases. 

\subsection{Unit Entropy}

Unit entropy measures the amount of entropy per path in the simulation. This is calculated in Python by dividing the internal entropy by the flow rate. It measures unit entropy at the end of the simulation, so the final 200 points of the internal entropy data are averaged, as are the final 200 points of the flow rate data. The averaged final entropy is then divided by the averaged flow rate: $s_{f}/\phi$.

\subsection{Simulation parameters}

\textbf{Parameter Values and Settings}
Tables D1 to D4 show the simulation parameters. 
\begin{table}
\centering
    \begin{tabularx}{\columnwidth}{|l|X|X|} \hline 
        Parameter & Value & Description \\ \hline 
        ant-speed & 1 patch/tick & Constant speed \\ \hline 
        wiggle range & 50 degrees & random directional change, \newline from -25 to +25 \\ \hline 
        view-angle & 135 degrees & Angle of cone where ants can detect pheromone \\ \hline 
        ant-size & 2 patches & Radius of ants, affects radius of pheromone viewing cone \\ \hline
    \end{tabularx}
\vspace{5pt}
\caption{These properties affect the behavior of the ants.}
\label{tab:SimParam}
\end{table}

\begin{table}
    \centering
    \begin{tabularx}{\linewidth}{|l|X|X|} \hline
        Parameter & Value & Description \\ \hline 
        Diffusion rate & 0.7 & Rate at which pheromones diffuse \\ \hline 
        Evaporation rate & 0.06 & Rate at which pheromones evaporate \\ \hline 
        Initial pheromone & 30 units & Initial amount of pheromone deposited \\ \hline
    \end{tabularx}
    \vspace{5pt}
    \caption{These properties affect the behavior of the pheromone.}
    \label{tab:Pheromone}
\end{table}
\begin{table}
    \centering
    \begin{tabularx}{\columnwidth}{|l|X|X|} \hline
        Parameter & Value & Description \\ \hline
        projectile-motion & off & Ants have constant energy \\ \hline
        start-nest-only & off & Ants start randomly \\ \hline
        max-food & 0 & Food is infinite, food will disappear if this is > 0 \\ \hline
        constant-ants & on & Number of ants is constant \\ \hline
        world-size & 40 & World ranges from -20 to +20, note that the true world size is 41x41 \\ \hline
    \end{tabularx}
    \vspace{5pt}
    \caption{These settings affect various other conditions. Some of them significantly change the simulation and are important to note.}
    \label{tab:Properties}
\end{table}
 \begin{table}
    \centering
    \begin{tabularx}{\columnwidth}{|l|X|X|} \hline 
        Parameter & Value & Description \\ \hline 
        food-nest-size & 5 & The length and width of the food and nest boxes \\ \hline 
        foodx & -18 & The position of the food in the x-direction \\ \hline 
        foody & 0 & The position of the food in the y-direction \\ \hline 
        nestx & +18 & The position of the nest in the x-direction \\ \hline 
        nesty & 0 & The position of the nest in the y-direction \\ \hline
    \end{tabularx}
    \vspace{5pt}
    \caption{The food/nest are boxes centered vertically on the screen. They do not move during the simulation. Horizontally, the back edges are aligned with the edge of the screen. They have a size of 5x5. To create this, set the following properties listed in the table below, then press the "box-food-nest button".}
    \label{tab:my_table}
\end{table}
\textbf{Analysis Parameters}
All datasets are averages of 20 runs for each population. There is also a moving average of 50 applied after standard averaging.

\subsection{Simulation Tests}
We ran several tests to show that the simulation and analysis were working correctly. 

\subsection{World Size}
Checking how many patches the world contains for the current setting. Running a command in NetLogo that counts how many patches are in a world with a size of 40 prints a value of 1681, and $\sqrt{1681}=41$. This means that when the world ranges from -20 to +20, the center patch is included, making a total of 41 patches in each direction.

\subsection{Estimated Path Area}
We run a test to check how well the estimation of how much area the ants occupy. We observed the algorithm working in vertical direction, when the ants are randomly dispersed and when they are on the horizontal path. When the ants were disperse, the estimated width was 46.8, which is slightly above the real world size of 41. When the ants formed a path, the estimated width was 5.6, which is close to the observed, with only a few outliers. So, the function that estimates path width might be a few patches off, but this is the due to stochastic behavior when averaging the positions. If, however, we did not use averaging, then the outlier ants would have an undesirable impact on the estimated width, and make the measurement fluctuate much more. The methods for checking the width and length of the path are identical, and these are both used in calculating the area occupied by the ants, which is an important step in calculating entropy and density.

\section{Results}

In this section, we present the data for self-organization as measured by different parameters as an output of the agent-based modeling simulations. First we present time data for raw output, from which points were obtained for the power law graphs. We show the evolution of some of the quantities from the beginning to the end of the simulation. The phase transition from disorder initially to order can be seen. The last 200 points averaged which have been used for the power law figures can be observed. The number of ants in the runs varies from 70 to 200. The time in the simulation runs from 0 to 1000 ticks. 

\subsection{Time graphs}

\begin{figure}[H]
    \includegraphics[width=8cm]{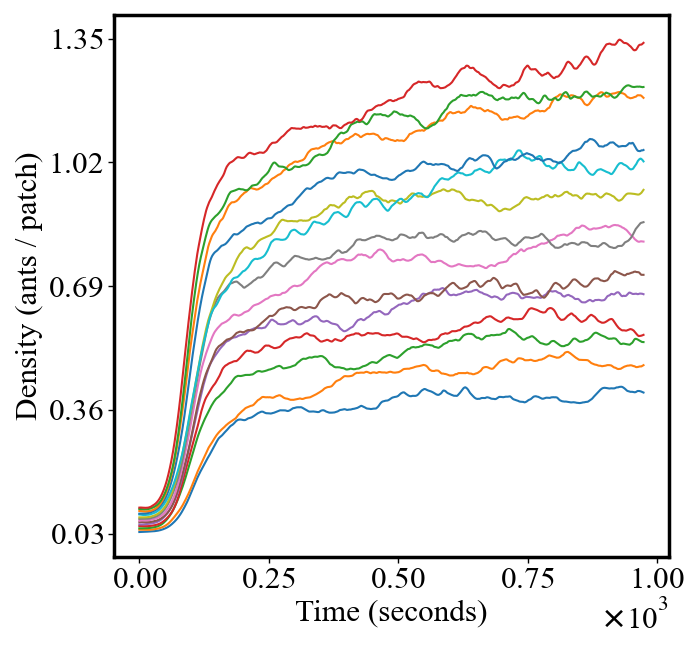}    \caption{The density of ants versus the time as the number of ants increases from the bottom curve to the top. As the simulation progresses, the ants become more dense.}
    \label{density}
\end{figure}

\begin{figure}[H]
    \includegraphics[width=8cm]{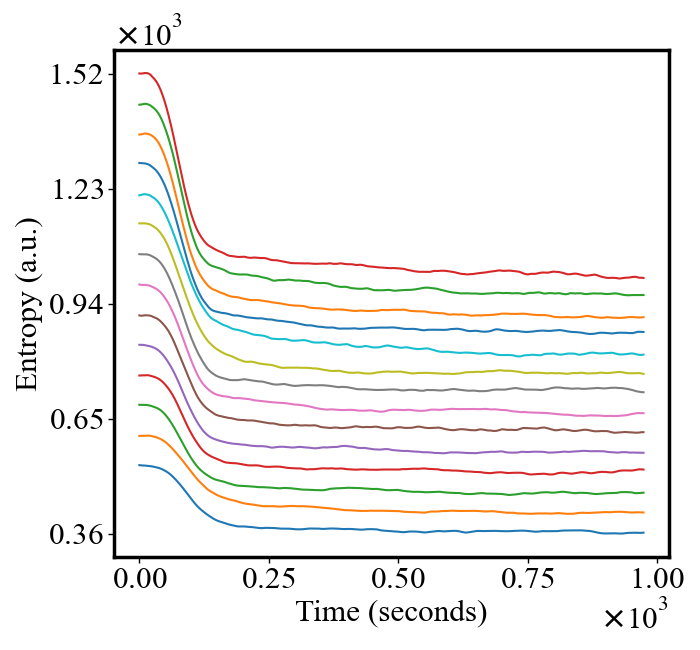}    \caption{The entropy of the simulation versus the time as the number of ants increases from bottom to top. Entropy decreases from the initial random state as the path forms.}
    \label{entropy}
\end{figure}

\begin{figure}[H]
    \includegraphics[width=8cm]{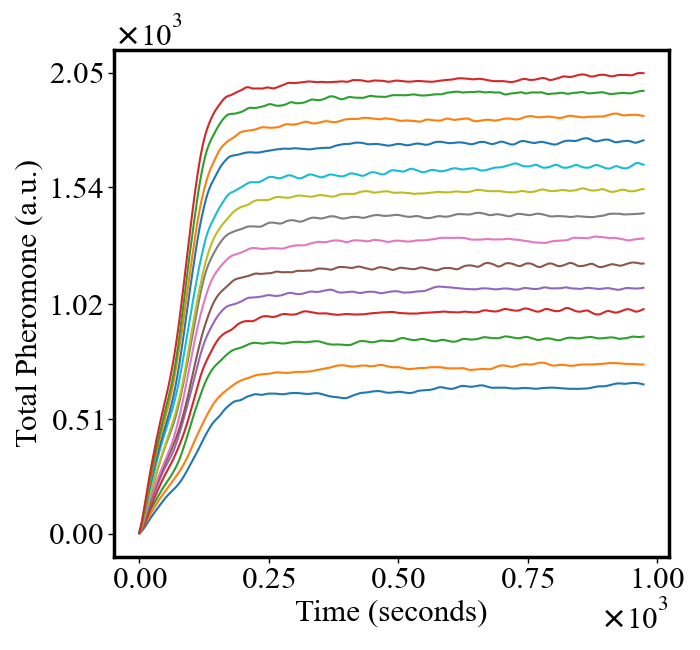}    \caption{The total amount of pheromone versus the time passed as the number of ants increases from bottom to top. As the simulation progresses, there is more pheromone for the ants to follow.}
    \label{pheromone}
\end{figure}

\begin{figure}[H]
    \includegraphics[width=8cm]{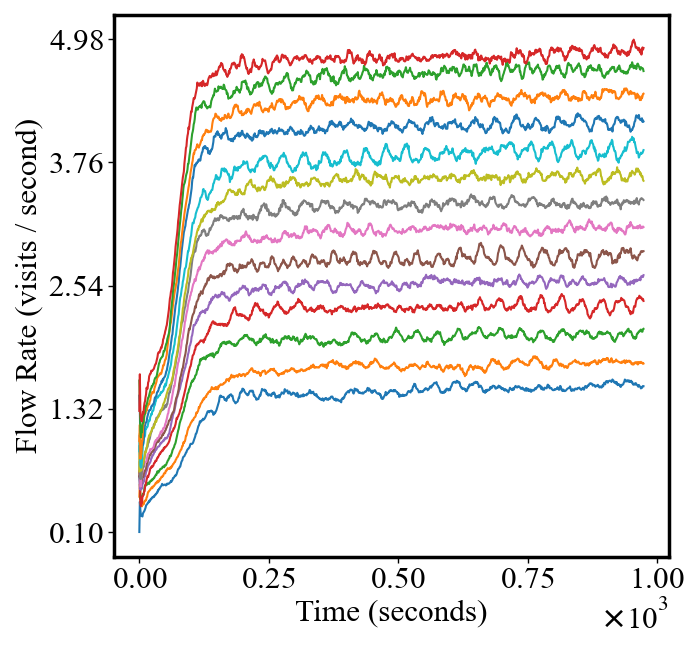}    \caption{The flow rate versus the time passed as the number of ants increases from bottom to top. As the simulation progresses, the ants visit the endpoints more often.}
    \label{flowRate}
\end{figure}

The raw data are presented as output measures vs time. In this paper we present the time graphs with the output for four quantities. \cref{density,entropy,pheromone,flowRate}.  All variables measure the degree of order in the system. 

The time data show the phase transition from a disorganized to an organized state as an increase of the order parameter \cref{density} and as a decrease in internal entropy \cref{entropy}. The amount of information \cref{pheromone} and the number of events per unit time \cref{flowRate} also undergo similar transitions. Those data are exponential in the region before the inflection point of the curves, where growth is unconstrained, confirming the exponential predictions of the model. 

The density serves as an order parameter \cref{density}, similar to entropy \cref{entropy}. In the first case, the system starts with a close to zero order parameter which increases to some maximum value and then saturates as the system is fixed in size. In the second case, the system starts at maximum internal entropy and it drops to a minimum value as the system reaches the saturation point. 

Pheromone is the amount of information in the system which is proportional to the degree of order \cref{pheromone}, and flow rate is also proportional to all of the other measures, indicating the number of events as defined in the system \cref{flowRate}. Both of these start at an initial minimum value and undergo a phase transition to the organized state, after which they saturate, due to the fixed size of the system. Those are measures directly connected to action efficiency and are some of the most important performance metrics for self-organizing systems.

The data here show the whole run as the density increases with time as self-organization occurs. The more ants, the larger the density. The increase of density depends on two factors: 1. The shorter average path length at the end, and 2. the increased number of ants.

The initial entropy when the ants are randomly dispersed scales with the number of ants, since the number of microstates corresponding to the same macrostate grows, and when they form the path, the amount of decrease of entropy is larger for the larger number of ants, even though the absolute value of entropy at the end of the simulation also scales with the number of ants, since the number of microstates when the ants form the final path also scales with the number of ants. Entropy at the initial and final state of the simulation is seen to grow with the increase of the number of agents.

The pheromone is a measure of the information in the system. It scales with the number of ants. Each simulation starts with zero pheromones as the ants are dispersed randomly and do not carry any pheromones, but as they start forming the path they lay pheromones from the food and nest respectively and the more ants are in the simulation, the more pheromones they carry. Larger systems contain more information as each agent is a carrier of information.
    
The number of events, which is the number of visits to the food and nest, is inversely proportional to the average path length, and respectively average path time, scales with the number of ants as expected. Initially, the number of crossings is zero but it quickly increases and is greater for the larger systems. After the ants form the shortest path, it saturates and stays close to constant for all simulations.

The average path length also can be used as a measure of the time and degree of self-organization. 

\subsection{Power law graphs}

All figures representing the relationship between the characteristics of this system demonstrate power law relationships between all of the quantities as theoretically predicted by the model. This serves as one confirmation of the model,  as a theoretical explanation for the simulation. The power law data correspond to scaling relations measured in many systems of different natures \citep{west2017scale, Bonner2004size}.

\subsubsection{Size-Complexity rule}

\begin{figure}[H]
    \includegraphics[width=8cm]{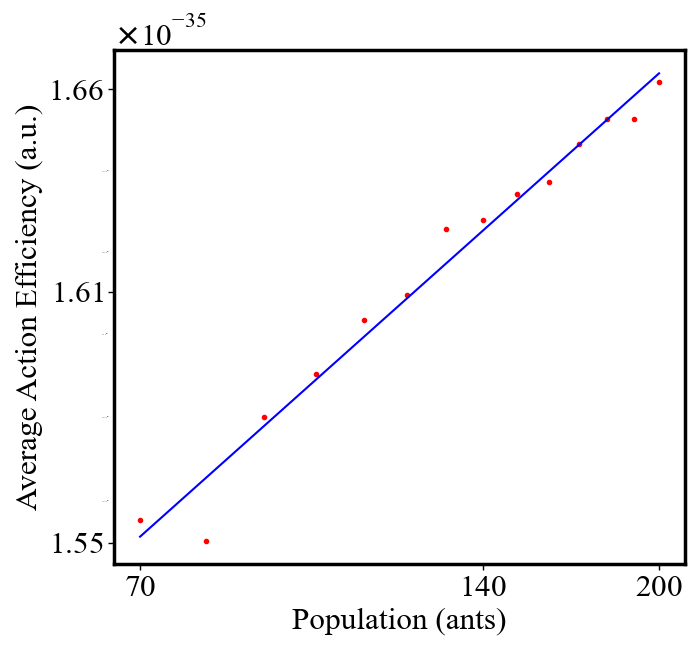}    \caption{The average action efficiency at the end of the simulation versus the number of ants. As more ants are added, they are able to form more action-efficient structures by finding shorter paths.}
    \label{a-N}
\end{figure}

\begin{figure}[H]
    \includegraphics[width=8cm]{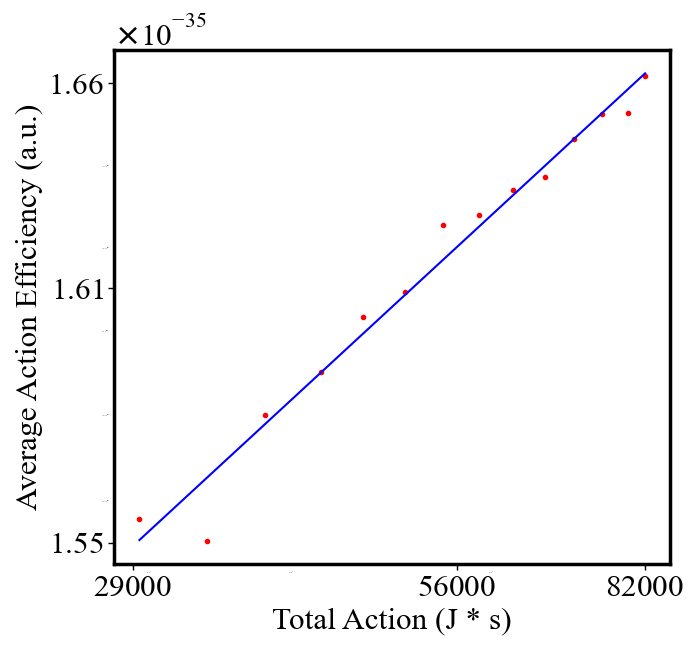}    \caption{The average action efficiency at the end of the simulation versus the total action as the number of ants increases. As there is more total action within the system, the ants become more action-efficient.}
    \label{a-Q}
\end{figure}

\begin{figure}[H]
    \includegraphics[width=8cm]{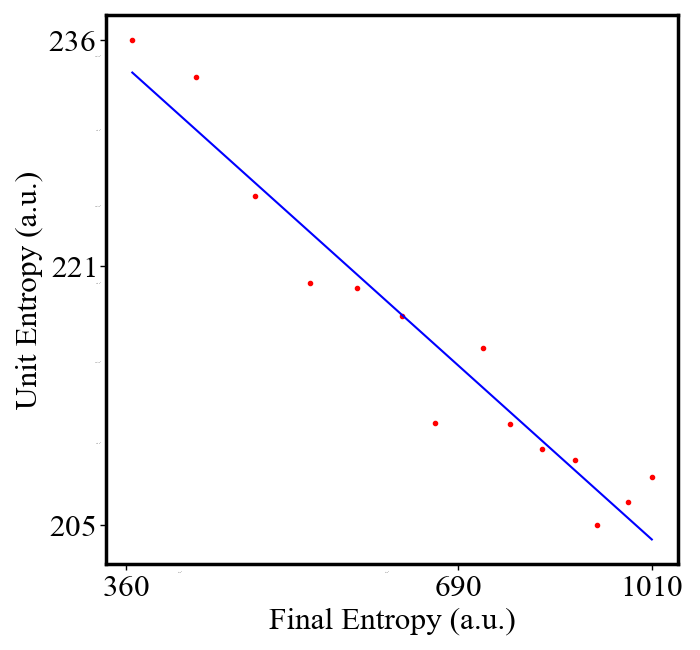}    \caption{Unit entropy at the end of the simulation versus internal entropy. As the total entropy for the simulation increases, the entropy per agent decreases.}
    \label{UnitEntVsFinEnt}
\end{figure}

Figure \ref{a-N} shows the size-complexity rule: as the size increases, the action efficiency as a measure of the degree of organization and complexity increases. This is supported by all experimental and observational data on scaling relations by Geoffrey West, Bonner, Carneiro, and many others \citep{Bonner2004size, west2017scale, carneiro1967relationship, Georgiev2021Stars}. They confirm Kleiber's law \citep{kleiber1932body}, and other similar laws, such as the area speciation rule in ecology and others.

\subsubsection{Unit-total dualism}

The following graphs serve as empirical support for the unit-total dualism described in this paper. 

The total action is a measure of all energy and time spent in the simulation by the agents in the system. As the number of agents increases, the total action increases. This demonstrates the duality of decreasing unit action and increasing total action as a system self-organizes grows, develops, and evolves. It also demonstrates the dynamical action principle as the unit action per one event decreases with the growth of the system, as seen in the increase of the average action efficiency, while the total action increases. This is an expression of the dualism for the decreasing unit action principle and the increasing total action principle, for dynamical action as systems self-organize, grow, evolve, and develop. 
This is an expression of the unit-total duality of entropy. When the unit entropy in the system tends to decrease its total entropy increases. 

\subsubsection{The rest of the characteristics}

Next we show the rest of the power law fits between all of the quantities in the model. All of them are on a log-log scale, where a straight line is a power law curve on a linear-linear scale. These graphs match the predictions of the model and confirm the power-law relationships between all of the characteristics of a complex system derived there. 

\begin{figure}[H]
    \includegraphics[width=8cm]{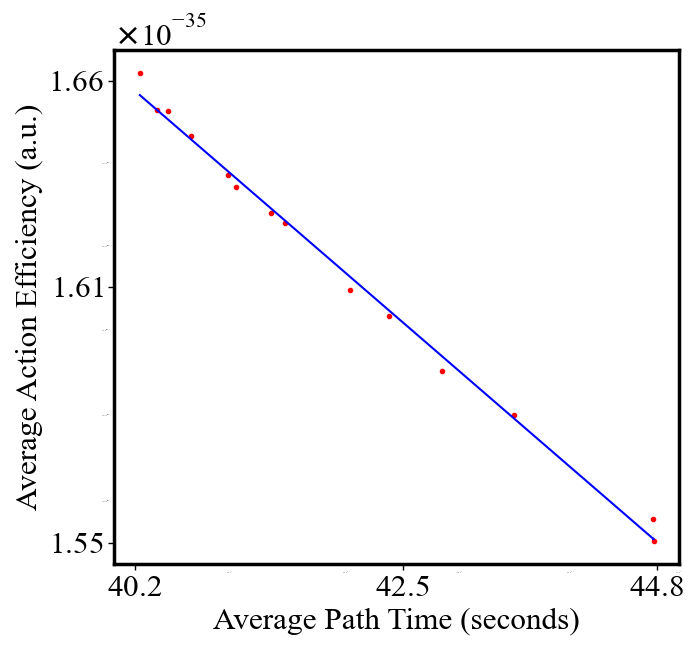}    \caption{The average action efficiency at the end of the simulation versus the time required to traverse the path as the number of ants increases. Action efficiency increases as the time to reach the destination shortens, i.e. the path length becomes shorter.}
    \label{a-A}
\end{figure}

In complex systems, as the agents find shorter paths, this state is more stable in dynamic equilibrium and is preserved. It has a higher probability of persisting. It is memorized by the system. If there is friction in the system, this trend will become even stronger, as the energy spent to traverse the shorter path will also decrease. To the macro-state at each point, there are many micro-states, corresponding to the variations of the paths of individual agents.

\begin{figure}[H]
    \includegraphics[width=8cm]{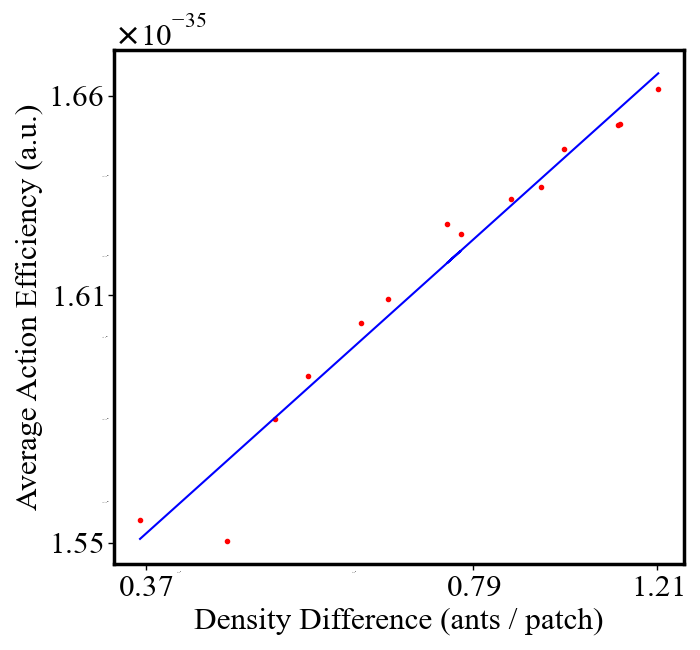}    \caption{The average action efficiency at the end of the simulation versus the density increase measured as the difference between the final density minus the initial density as the number of ants increases. As the ants get more dense, they become more action efficient.}
    \label{a-p}
\end{figure}

Density increases the probability of shorter paths, i.e. less time to reach the destination, i.e. larger action efficiency. In natural systems as density increases, action efficiency increases, i.e. level of organization increases. Another term for density is concentration. When hydrogen gas clouds in the universe under the influence of gravity concentrate into stars, nucleosynthesis starts and the evolution of cosmic elements begins. In chemistry increased concentration of reactants speeds up chemical reactions, i.e. they become more action efficient. When single-cell organisms concentrate in colonies and later in multicellular organisms their level of organization increases. When human populations concentrate in cities, action efficiency increases, and civilization advances.

\begin{figure}[H]
    \includegraphics[width=8cm]{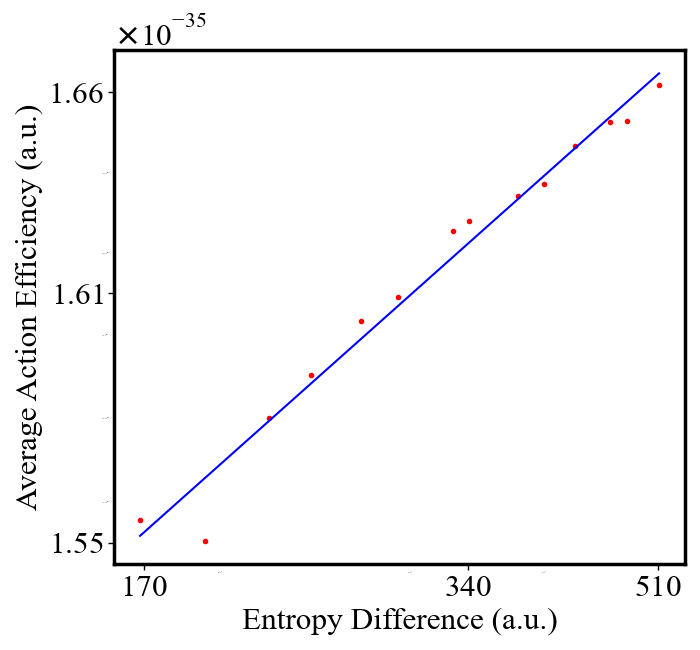}    \caption{The average action efficiency at the end of the simulation versus the absolute amount of entropy decreases as the number of ants increases. As the ants get less random, they become more action-efficient. }
    \label{a-s}
\end{figure}

As statistical Boltzmann entropy decreases, the density (concentration) increases and this allows the system to be more action-efficient and organized. Decreased randomness is correlated with a well-formed path as a flow channel, which corresponds to the structure (organization) of the system. Here, the decrease of entropy obeys the predictions of the model being in a strict power law dependence on the other characteristics of the self-organizing complex system.

\begin{figure}[H]
    \includegraphics[width=8cm]{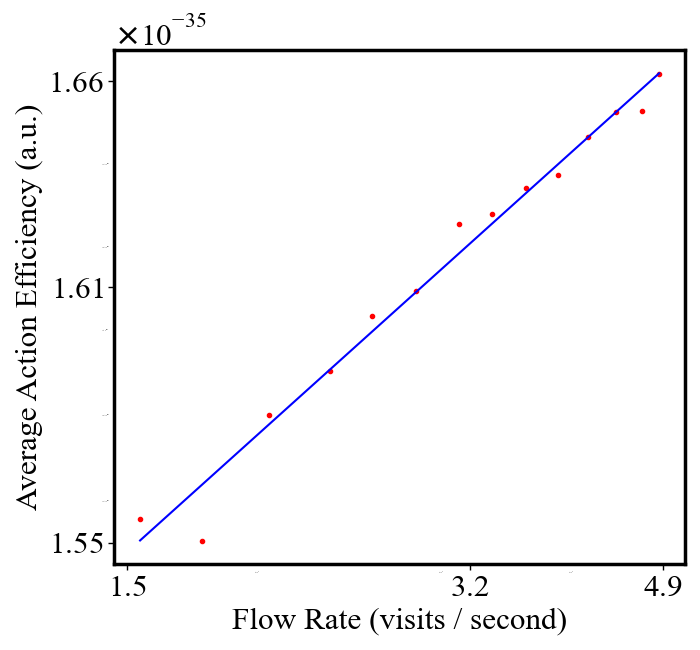}    \caption{The average action efficiency at the end of the simulation versus the flow rate as the number of ants increases. As the ants visit the endpoints more often, they become more efficient.}
    \label{a-f}
\end{figure}
The flow rate measures the number of events in a system. For real systems, those can be nuclear or chemical reactions, computations, or anything else. In this simulation, it is the number of visits at the endpoints, or the number of crossings. As the speed of the ants is a constant in this simulation, the number of visits or the flow of events is inversely proportional to the time for crossing, i.e. the path length, therefore action efficiency increases with the number of visits.

\begin{figure}[H]
    \includegraphics[width=8cm]{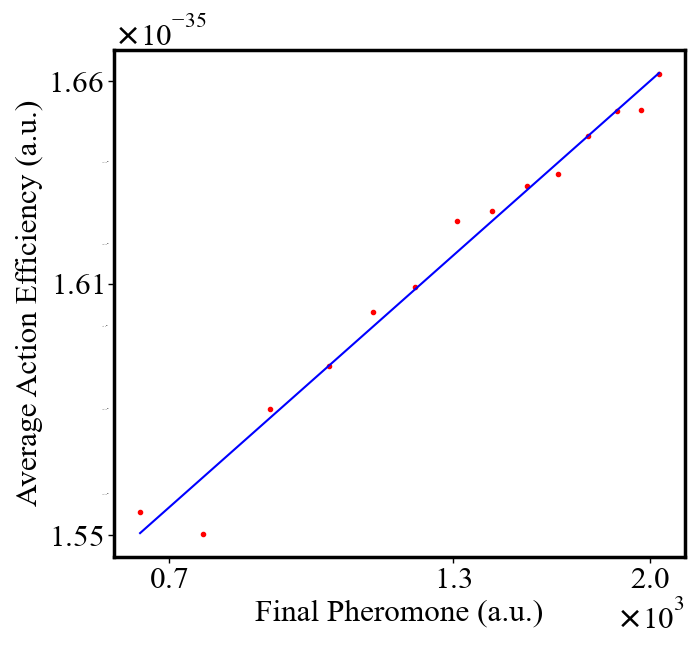}    \caption{The average action efficiency at the end of the simulation versus the amount of pheromone, or information, as the number of ants increases. As there is more information for the ants to follow, they become more efficient.}
    \label{a-i}
\end{figure}

The pheromone is what instructs the ants how to move. They follow its gradient towards the food or the nest. As the ants form the path, they concentrate more pheromone on the trail, and they lay it faster so it has less time to evaporate. Both depend on each other in a positive feedback loop. This leads to increased action efficiency, with a power-law dependence as predicted by the model. In other complex systems, the analog of the pheromone can be temperature and catalysts in chemical reactions. In an ecosystem, as animals traverse a path, the path itself carries information, and clearing the path reduces obstacles and, therefore the time and energy to reach the destination, i.e. action.

\begin{figure}[H]
    \includegraphics[width=8cm]{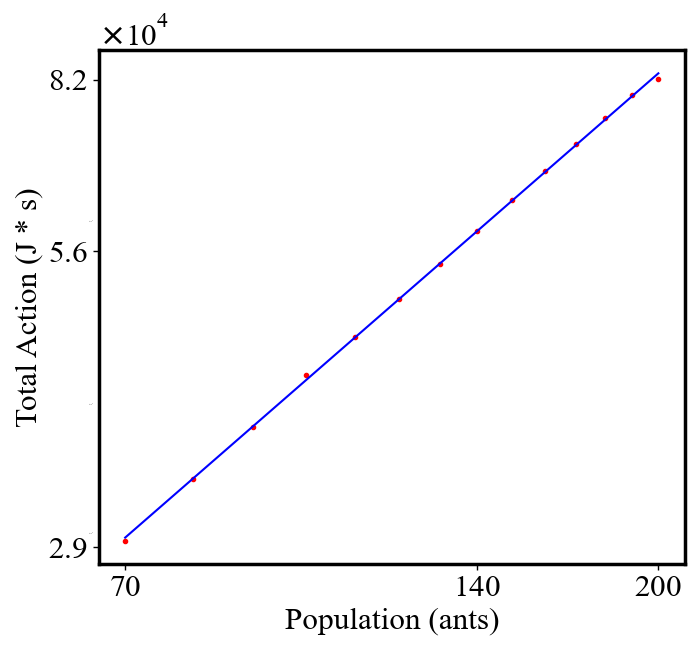}    \caption{The total action at the end of the simulation versus the number of ants. As more ants are added, there is action energy within the system.}
    \label{Q-n}
\end{figure}

The total action is the sum of the actions of each agent. As the number of agents grows the total action grows.  This graph demonstrates the principle of increasing total action in self-organization, growth, evolution, and development of systems. 

\begin{figure}[H]
    \includegraphics[width=8cm]{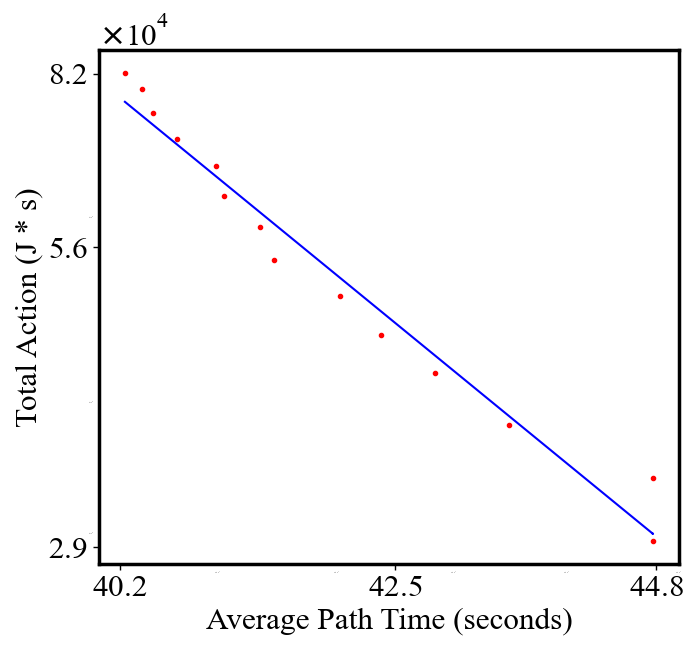}    \caption{The total action at the end of the simulation versus the time required to traverse the path as the number of ants increases.}
    \label{Q-A}
\end{figure}

With more ants, the path forms better and gets shorter, which increases the number of visits. The shorter time is connected to more visits and increased size of the system, which is why the total action increases. This graph also demonstrates the principle of increasing total action in self-organization, growth, evolution, and development of systems. 

\begin{figure}[H]
    \includegraphics[width=8cm]{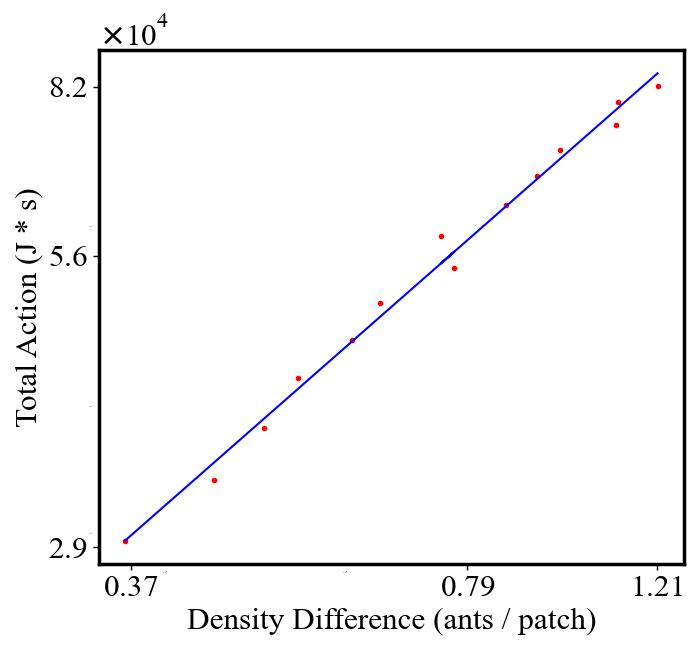}    \caption{The total action at the end of the simulation versus the increase of density as the number of ants increases. As the ants become more dense, there is more action in the system.}
    \label{Q-p}
\end{figure}

The larger the system is, it contains more agents, which corresponds to greater density, more trajectories, and more total action. This graph demonstrates as well the principle of increasing total action in self-organization, growth, evolution, and development of systems. 

\begin{figure}[H]
    \includegraphics[width=8cm]{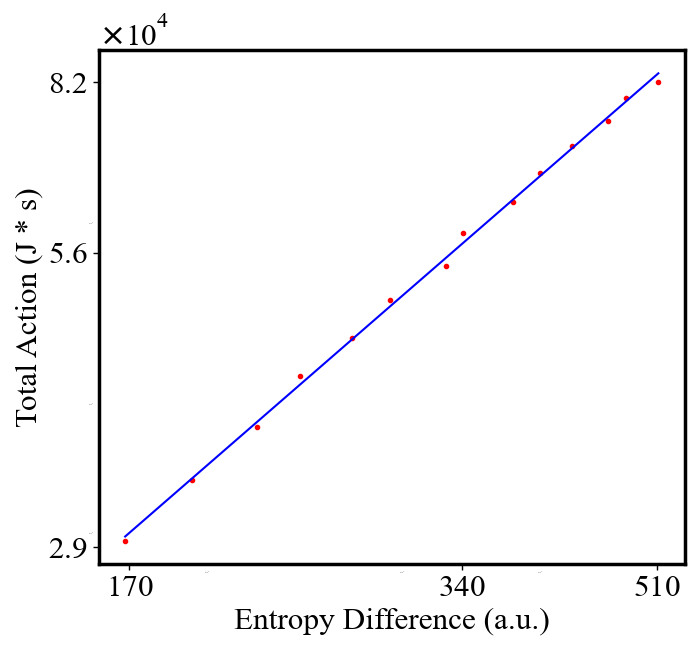}    \caption{The total action at the end of the simulation versus the absolute decrease of entropy as the number of ants increases. As the entropy decreases, there is more action within the system.}
    \label{Q-s}
\end{figure}

As the total entropy difference increases, which means that the decrease of the internal entropy is greater for a larger number of ants, the total action increases, because there are more agents in the system and they visit the nodes more often. Greater organization of the system is correlated with more total action demonstrating again the principle of increasing total action in self-organization, growth, evolution, and development of systems. 

\begin{figure}[H]
    \includegraphics[width=8cm]{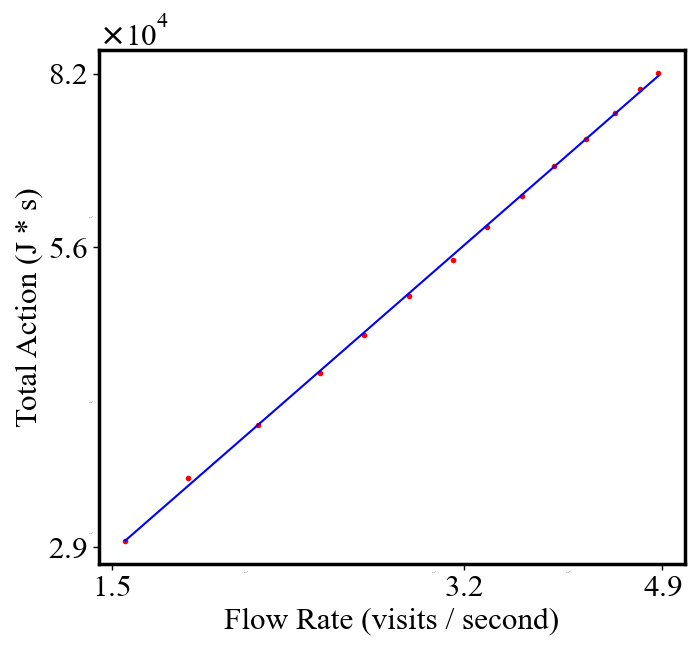}    \caption{The total action at the end of the simulation versus the flow rate as the number of ants increases. As the ants visit the endpoints more often, there is more total action within the system.}
    \label{Q-f}
\end{figure}

As the flow of events increases, which is the number of crossings of ants between the food and nest, the total action increases, because there are more agents in the system and they visit the nodes more often by forming a shorter path. This also demonstrates the principle of increasing total action in self-organization, growth, evolution, and development of systems. 

\begin{figure}[H]
    \includegraphics[width=8cm]{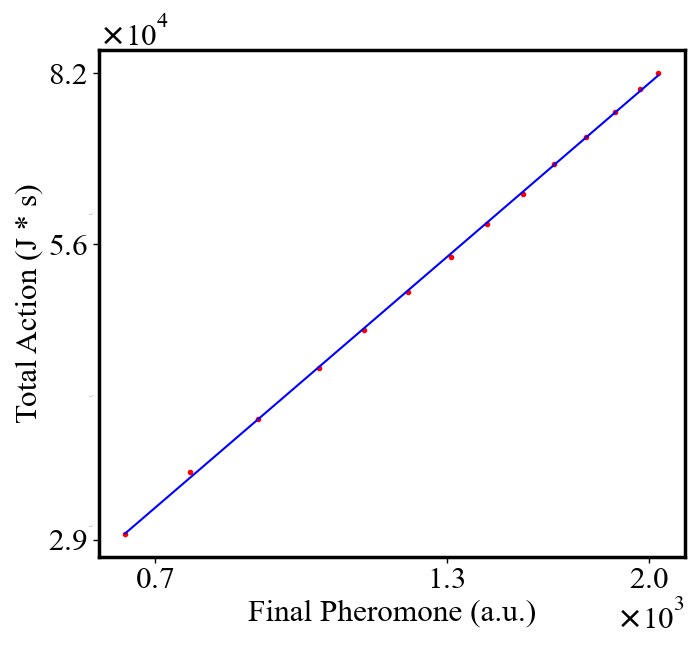}    \caption{The total action at the end of the simulation versus the amount of pheromone as the number of ants increases. As there is more information for the ants to follow, there is more action within the system.}
    \label{Q-i}
\end{figure}

As the total number of agents in the system increases, they leave more pheromones, which causes forming a shorter path, increases the number of visits, and the total action increases. Again, this graph demonstrates the principle of increasing total action in self-organization, growth, evolution, and development of systems. 

\begin{figure}[H]
    \includegraphics[width=8cm]{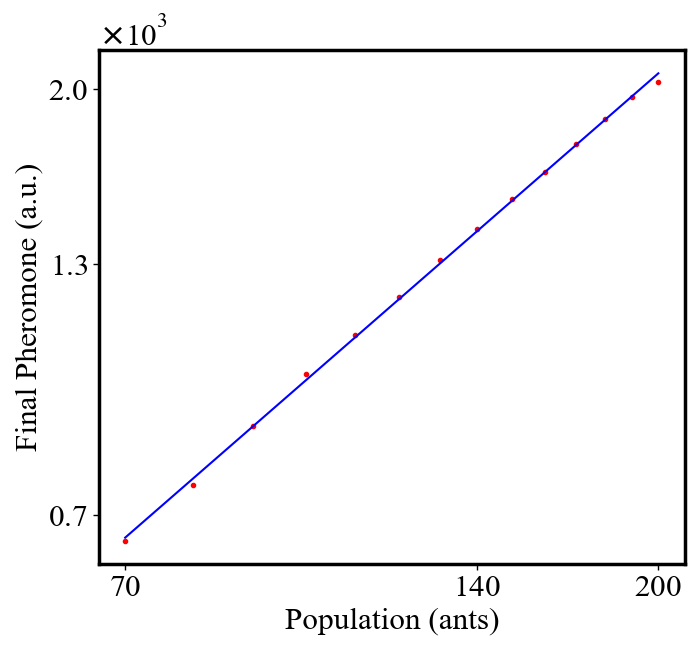}    \caption{The total pheromone at the end of the simulation versus the number of ants. As more ants are added to the simulation, there is more information for the ants to follow.}
    \label{i-N}
\end{figure}

As the total number of ants in the system increases, they leave more pheromones and form a shorter path, which counters the evaporation of the pheromones. This increases the amount of information in the system, which helps with its rate and degree of self-organization.

\begin{figure}[H]
    \includegraphics[width=8cm]{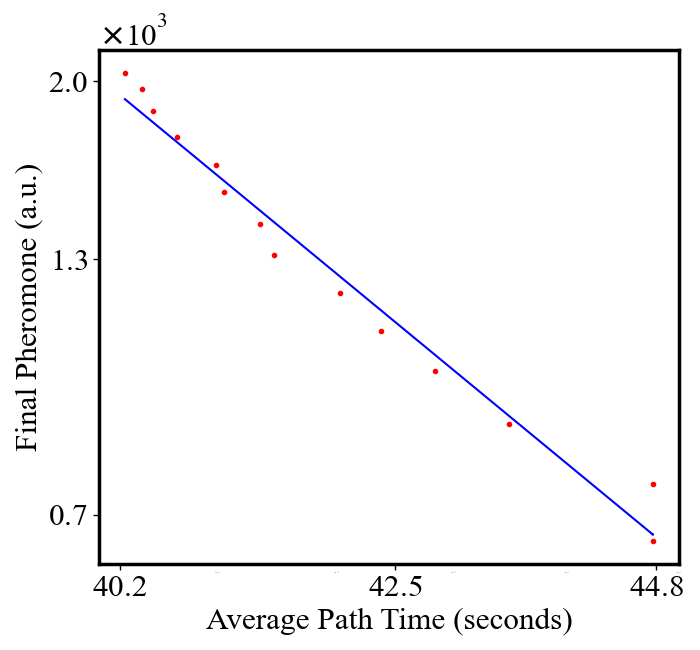}    \caption{The total pheromone at the end of the simulation versus the time required to traverse the path as the number of ants increases. As it takes less time for the ants to travel between the nodes, there is more information for the ants to follow. }
    \label{i-A}
\end{figure}

As the total number of ants in the system increases, they form a shorter path as the degree of self-organization is higher, and as they visit the food and nest more often and there are greater number of ants they leave more pheromones. The increased amount of information in turn helps form an even shorter path which reduces the pheromone evaporation increasing the pheromones event more. This is a visualization of the result of this positive feedback loop.

\begin{figure}[H]
    \includegraphics[width=8cm]{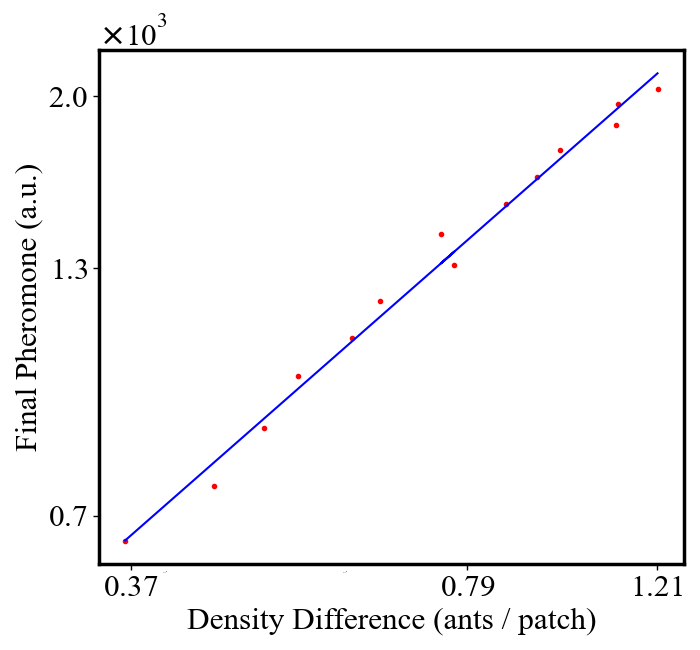}    \caption{The total pheromone at the end of the simulation versus the density increase as the number of ants increases. As the ants become more dense, there is more information for them to follow.}
    \label{i-p}
\end{figure}

As the total number of ants in the system increases, they form a shorter path as the degree of self-organization is higher, and as there are more ants, their density increases, and as they visit the food and nest more often and there is a greater number of ants and lower evaporation, they leave more information.

\begin{figure}[H]
    \includegraphics[width=8cm]{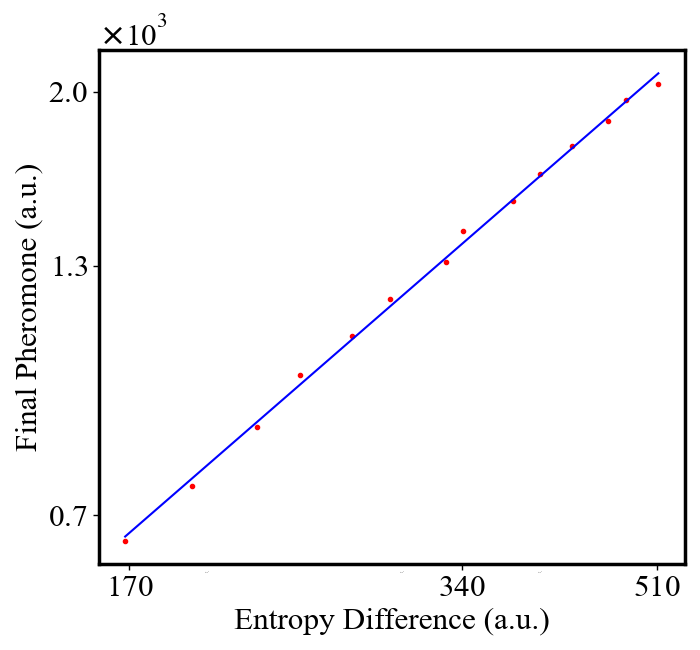}    \caption{The total pheromone at the end of the simulation versus the absolute decrease of entropy as the number of ants increases. As the entropy decreases, there is more information for the ants to follow.}
    \label{i-s}
\end{figure}

As the total number of ants in the system increases, they form a shorter path as the degree of self-organization is higher.  As there are more ants, the entropy difference increases. The entropy during each simulation decreases, and as they visit the food and nest more often and there is a greater number of ants and less evaporation they accumulate more pheromones. 

\begin{figure}[H]
    \includegraphics[width=8cm]{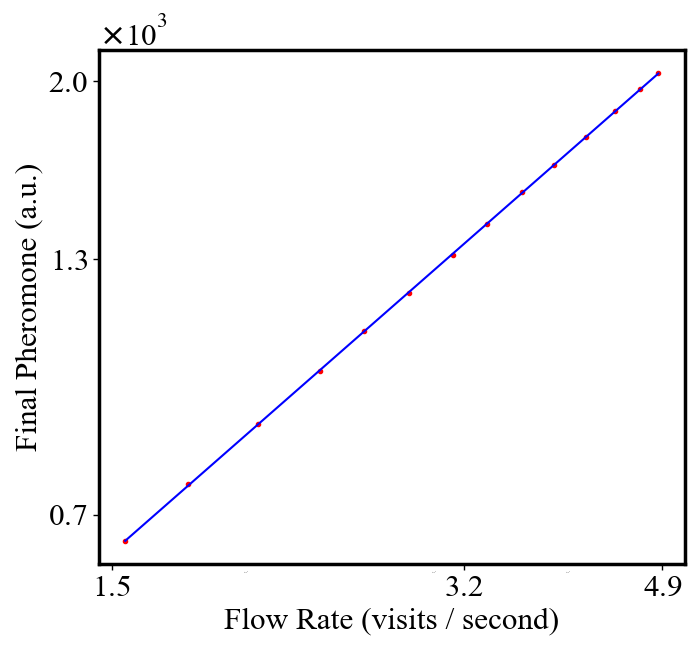}    \caption{The total pheromone at the end of the simulation versus the flow rate as the number of ants increases. }
    \label{i-f}
\end{figure}

As the total number of ants in the system increases, they form a shorter path as the degree of self-organization is higher. They visit the food and nest more often, and as there are more ants, the number of visits increases proportionally, the evaporation decreases, and they accumulate more pheromones. 

\begin{figure}[H]
    \includegraphics[width=8cm]{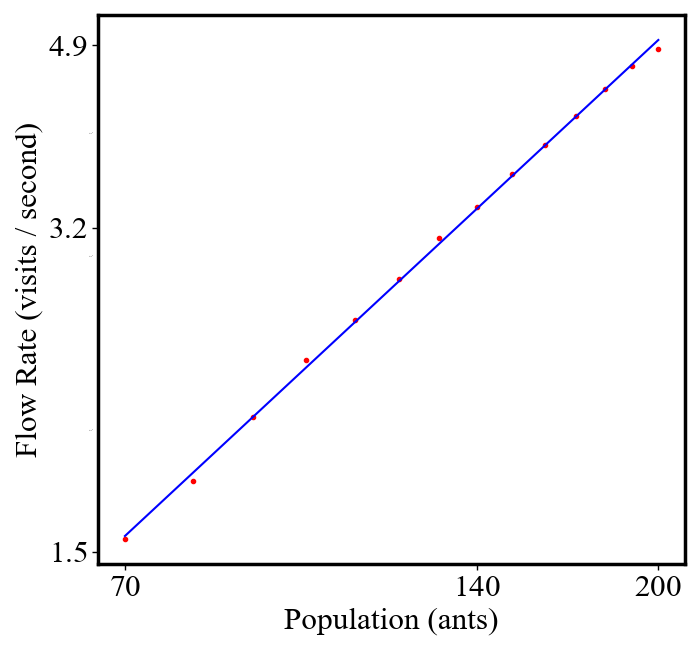}    \caption{The flow rate at the end of the simulation versus the number of ants. As more ants are added to the simulation and they are forming shorter paths in self-organization, the ants are visiting the endpoints more often.}
    \label{f-N}
\end{figure}

As the total number of ants in the system increases, they form a shorter path as the degree of self-organization is higher, visit the food and nest more often, and the number of visits increases proportionally. 

\begin{figure}[H]
    \includegraphics[width=8cm]{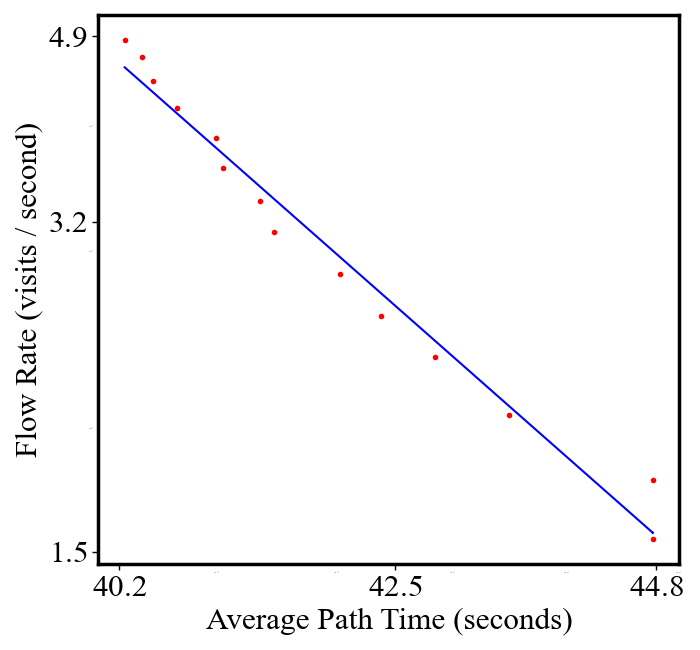}    \caption{The flow rate at the end of the simulation versus the time required to traverse between the nodes as the number of ants increases. As the path becomes shorter, the ants are visiting the endpoints more often.}
    \label{f-A}
\end{figure}

As the total number of ants in the system increases, they form a shorter path as the degree of self-organization is higher, visit the food and nest more often, and as there are more ants, the number of visits increases proportionally. 

\begin{figure}[H]
    \includegraphics[width=8cm]{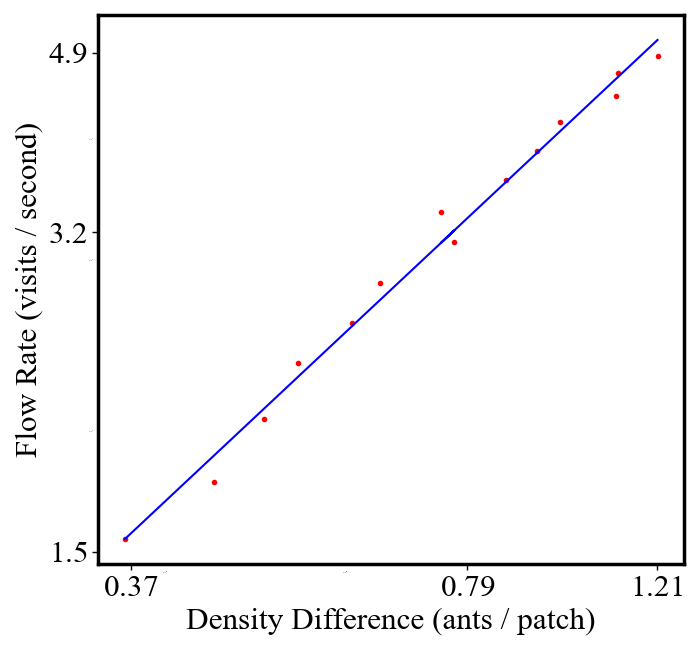}    \caption{The flow rate at the end of the simulation versus the increase of density as the number of ants increases. As the ants get more dense, they are visiting the endpoints more often.}
    \label{f-p}
\end{figure}

As the total number of ants in the system increases, they form a shorter path as the degree of self-organization is higher, this leads to an increase in density, and as there are more ants, the number of visits increases proportionally. 

\begin{figure}[H]
    \includegraphics[width=8cm]{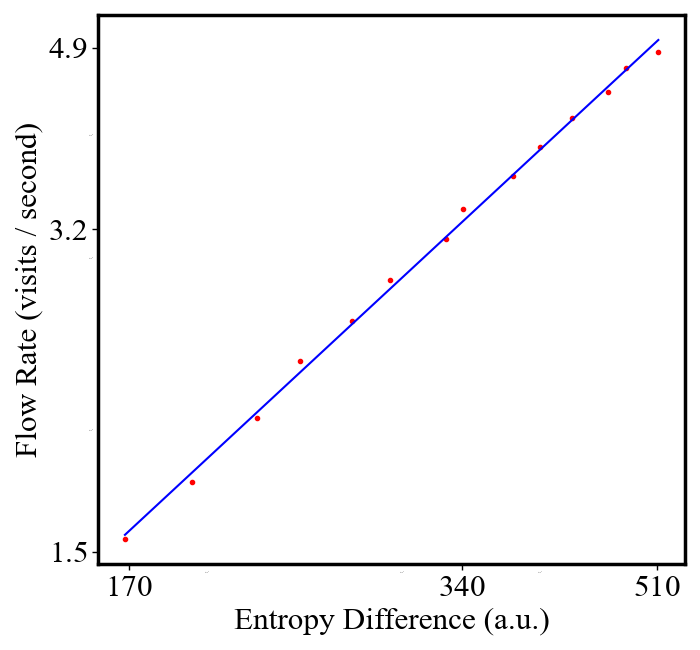}    \caption{The flow rate at the end of the simulation versus the absolute decrease of entropy as the number of ants increases. As the entropy decreases more, the ants are visiting the endpoints more often.}
    \label{f-s}
\end{figure}

As the total number of ants in the system increases, they form a shorter path as the degree of self-organization is higher, the absolute decrease of entropy is larger, and as there are more ants, the number of visits increases proportionally. 

\begin{figure}[H]
    \includegraphics[width=8cm]{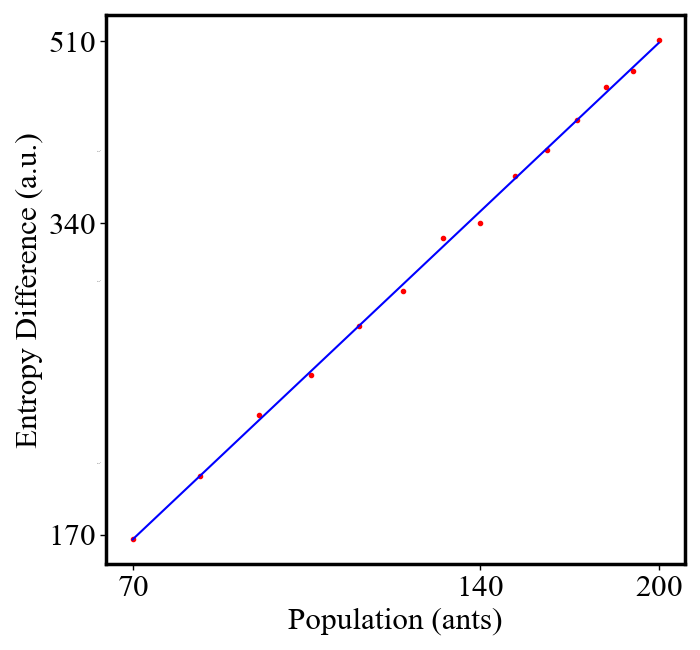}    \caption{The absolute amount of entropy decrease versus the number of ants. As more ants are added to the simulation, there is a larger decrease in entropy reflecting a greater degree of self-organization.}
    \label{s-N}
\end{figure}

As the total number of ants in the system increases, they form a shorter path as the degree of self-organization is higher, they start with a larger initial entropy and the difference between the initial and final entropy grows. More ants correspond to greater internal entropy decrease, which is one measure of self-organization. It is one of the scaling laws in the size-complexity rule. 

\begin{figure}[H]
    \includegraphics[width=8cm]{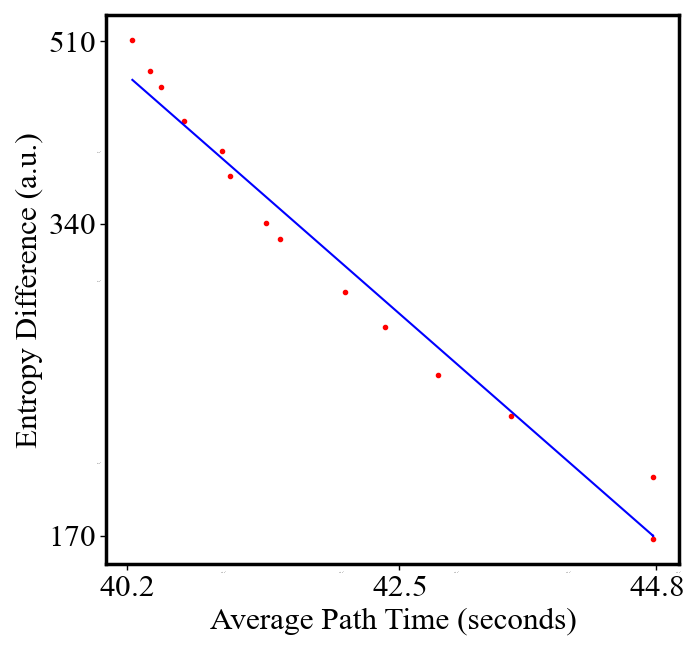}    \caption{The absolute amount of entropy decrease versus the time required to traverse the path at the end of the simulation as the number of ants increases. As it takes more time to move between the nodes with fewer ants, there is less of a decrease in entropy.}
    \label{s-A}
\end{figure}

As the total number of ants in the system increases, they form a shorter path as the degree of self-organization is higher. When the path is shorter, this corresponds to shorter times to cross between the two nodes, the internal entropy decreases more. 

\begin{figure}[H]
    \includegraphics[width=8cm]{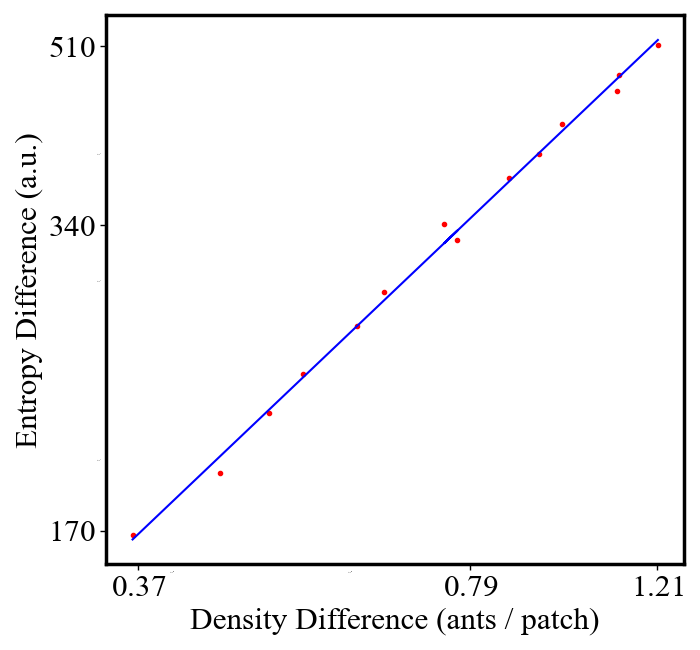}    \caption{The absolute amount of entropy decrease versus the amount of density increase as the number of ants increases. As the ants become more dense, there is a larger decrease in entropy.}
    \label{s-p}
\end{figure}

As the total number of ants in the system increases, they form a shorter path as the degree of self-organization is higher, and as there are more ants their density increases, and the internal entropy difference increases proportionally. 

\begin{figure}[H]
    \includegraphics[width=8cm]{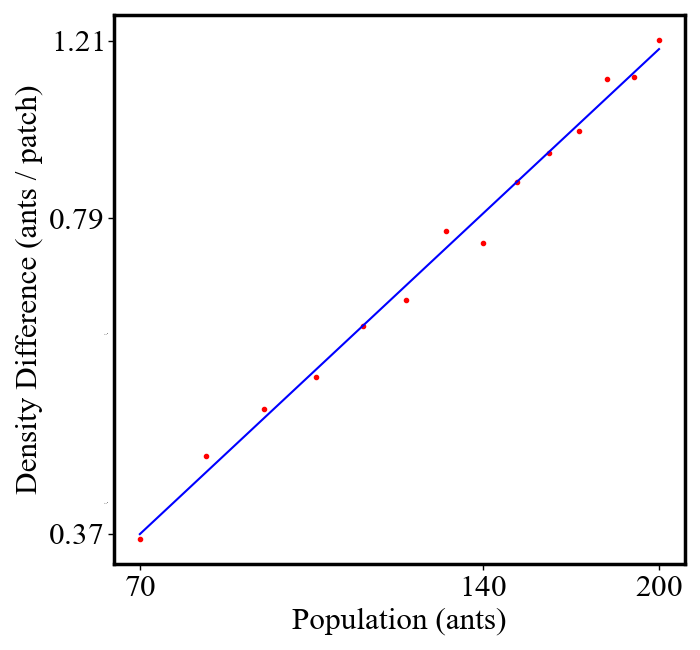}    \caption{The amount of density increase versus the number of ants. As more ants are added to the simulation, there is a larger increase in density.}
    \label{p-N}
\end{figure}

As the total number of ants in the system increases, they form a shorter path as the degree of self-organization is higher, and as there are more ants, the density increases proportionally. 

\begin{figure}[H]
    \includegraphics[width=8cm]{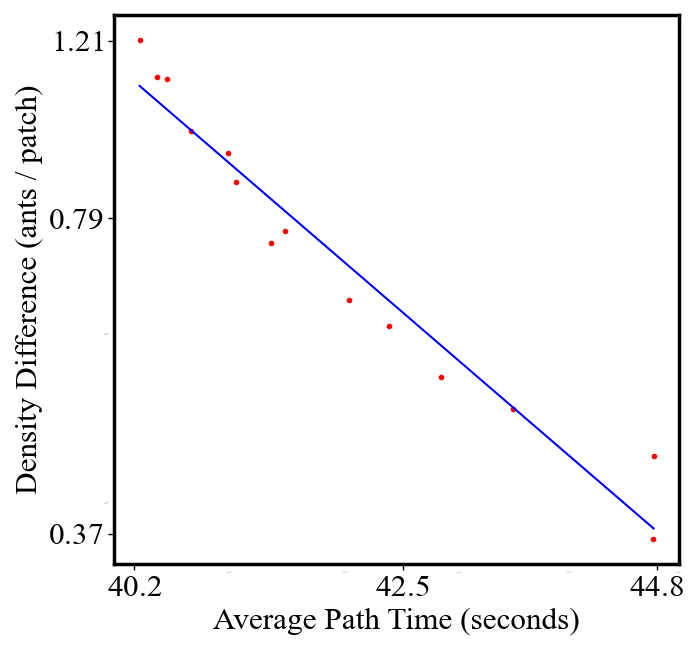}    \caption{The amount of density increase versus the time required to traverse the path as the number of ants increases. When there are more ants it takes less time to traverse the path, and there is more of an increase in density.}
    \label{p-A}
\end{figure}

As the total number of ants in the system increases, they form a shorter path as the degree of self-organization is higher, visit the food and nest more often, the time to cross between the nodes decreases, and the density increases proportionally. 

\begin{figure}[H]
    \includegraphics[width=8cm]{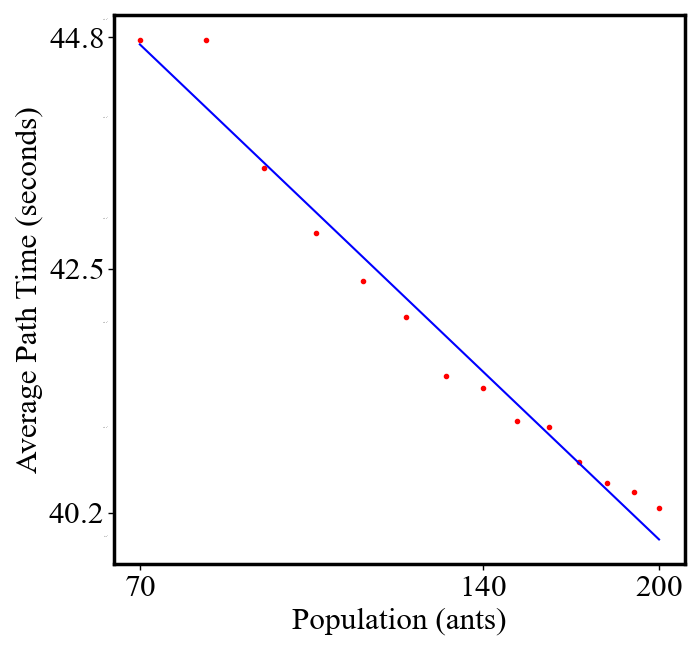}    \caption{The time required to traverse the path versus the number of ants. As more ants are added to the simulation, it takes less time to move between the nodes.}
    \label{A-N}
\end{figure}

As the total number of ants in the system increases, they form a shorter path as the degree of self-organization is higher, visit the food and nest more often, and as there are more ants, the time for the visits decreases proportionally.

\begin{figure}[H]
    \includegraphics[width=8cm]{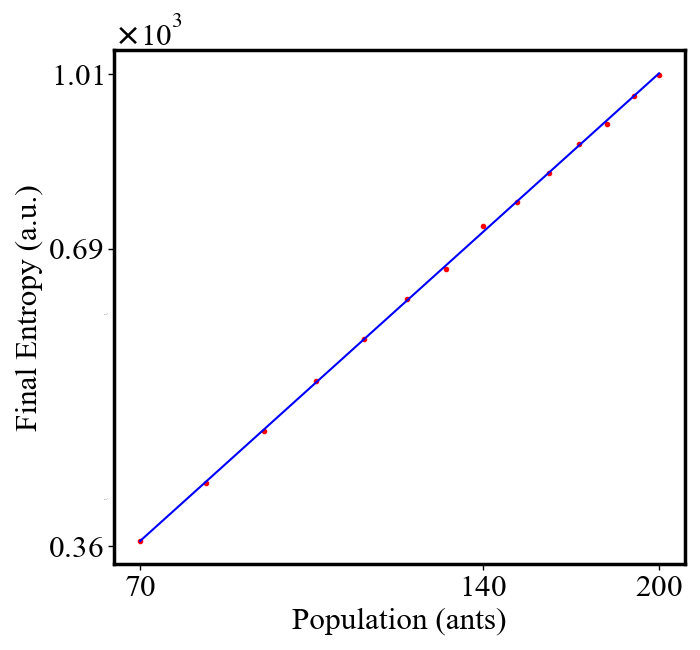}    \caption{Population versus the average entropy at the end of the simulation. As the population increases, there is more entropy.}
    \label{FinEnt}
\end{figure}
The final entropy in the system increases when there are more agents, and therefore more possible microstates of the system. This is an expression of the unit-total duality of entropy when the total entropy in the system tends to increase with its growth. 
\begin{figure}[H]
    \includegraphics[width=8cm]{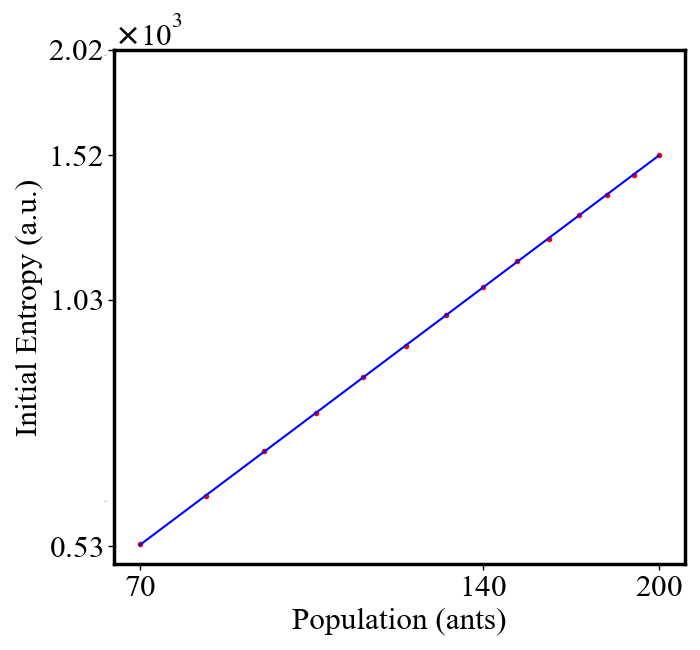}    \caption{Population versus the initial entropy on the first tick of the simulation. As the population increases, there is more entropy.}
    \label{InitEnt}
\end{figure}
The initial entropy reflects the larger number of agents in a fixed initial size of the system and scales with the size of the system as expected. The initial entropy in the system increases when there are more agents in the space of the simulation, and therefore more possible microstates of the system.
\begin{figure}[H]
    \includegraphics[width=8cm]{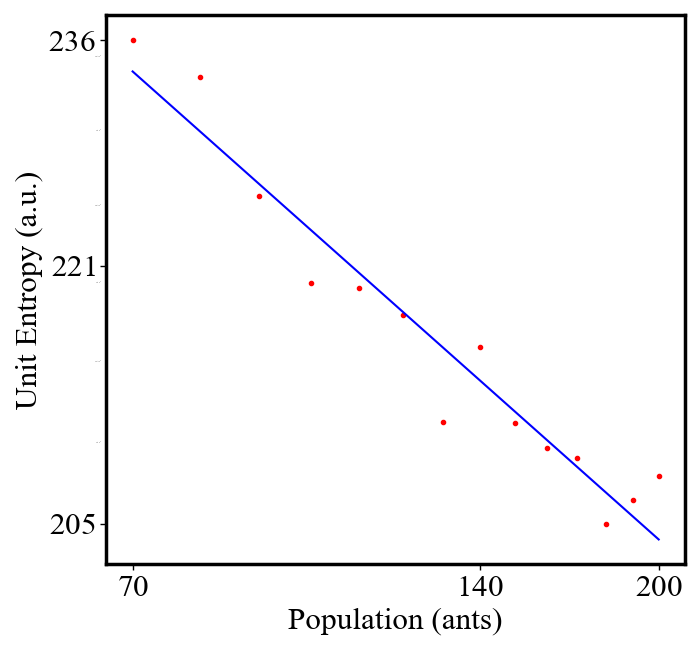}    \caption{Unit entropy at the end of the simulation versus population. As there are more agents, there is less entropy per path at the end of the simulation.}
    \label{UnitEntVsPop}
\end{figure}
This is an expression of the unit-total duality of entropy when the unit entropy in the system tends to decrease with its growth. 

In the next table we show the values of the fit parameters for the power law relationships

\captionsetup[table]{font={up}}

\begin{table}
    \centering
    \renewcommand{\arraystretch}{1.3}
    \begin{tabular}{|l|l|l|l|} \hline
        variables&a&b&$R^{2}$\\ \hline
        $\langle\alpha\rangle$\ vs.\ Q&$7.713\cdot 10^{-36}$&$6.787\cdot 10^{-2}$&$0.977$\\ \hline
        $\langle\alpha\rangle$\ vs.\ i&$1.042\cdot 10^{-35}$&$6.131\cdot 10^{-2}$&$0.981$\\ \hline
        $\langle\alpha\rangle$\ vs.\ $\phi$&$1.510\cdot 10^{-35}$&$6.055\cdot 10^{-2}$&$0.982$\\ \hline
        $\langle\alpha\rangle$\ vs.\ $\Delta s$&$1.020\cdot 10^{-35}$&$6.675\cdot 10^{-2}$&$0.978$\\ \hline
        $\langle\alpha\rangle$\ vs.\ $\Delta\rho$&$1.647\cdot 10^{-35}$&$5.947\cdot 10^{-2}$&$0.964$\\ \hline
        $\langle\alpha\rangle$\ vs.\ $\langle t \rangle$&$1.622\cdot 10^{-34}$&$-6.175\cdot 10^{-1}$&$0.995$\\ \hline
        $\langle\alpha\rangle$\ vs.\ N&$1.168\cdot 10^{-35}$&$6.673\cdot 10^{-2}$&$0.977$\\ \hline
        Q\ vs.\ i&$8.502\cdot 10^{1}$&$9.012\cdot 10^{-1}$&$1.000$\\ \hline
        Q\ vs.\ $\phi$&$2.000\cdot 10^{4}$&$8.897\cdot 10^{-1}$&$1.000$\\ \hline
        Q\ vs.\ $\Delta s$&$6.202\cdot 10^{1}$&$9.829\cdot 10^{-1}$&$0.999$\\ \hline
        Q\ vs.\ $\Delta\rho$&$7.133\cdot 10^{4}$&$8.784\cdot 10^{-1}$&$0.990$\\ \hline
        Q\ vs.\ $\langle t \rangle$&$1.410\cdot 10^{19}$&$-8.888$&$0.972$\\ \hline
        Q\ vs.\ N&$4.550\cdot 10^{2}$&$9.830\cdot 10^{-1}$&$1.000$\\ \hline
        i\ vs.\ $\phi$&$4.281\cdot 10^{2}$&$9.873\cdot 10^{-1}$&$1.000$\\ \hline
        i\ vs.\ $\Delta s$&$7.064\cdot 10^{-1}$&$1.090$&$0.999$\\ \hline
        i\ vs.\ $\Delta\rho$&$1.755\cdot 10^{3}$&$9.740\cdot 10^{-1}$&$0.988$\\ \hline
        i\ vs.\ $\langle t \rangle$&$1.407\cdot 10^{19}$&$-9.887$&$0.976$\\ \hline
        i\ vs.\ N&$6.445$&$1.090$&$0.999$\\ \hline
        $\phi$\ vs.\ $\Delta s$&$1.521\cdot 10^{-3}$&$1.104$&$0.999$\\ \hline
        $\phi$\ vs.\ $\Delta\rho$&$4.175$&$9.864\cdot 10^{-1}$&$0.988$\\ \hline
        $\phi$\ vs.\ $\langle t \rangle$&$5.438\cdot 10^{16}$&$-1.002\cdot 10^{1}$&$0.977$\\ \hline
        $\phi$\ vs.\ N&$1.427\cdot 10^{-2}$&$1.104$&$0.999$\\ \hline
        $\Delta s$\ vs.\ $\Delta\rho$&$1.301\cdot 10^{3}$&$8.939\cdot 10^{-1}$&$0.991$\\ \hline
        $\Delta s$\ vs.\ $\langle t \rangle$&$4.439\cdot 10^{17}$&$-9.035$&$0.969$\\ \hline
        $\Delta s$\ vs.\ N&$7.598$&$1.000$&$1.000$\\ \hline
        $s_{i}$\ vs.\ N&$7.598$&$1.000$&$1.000$\\ \hline
        $s_{f}$\ vs.\ N&$5.793$&$9.745\cdot 10^{-1}$&$1.000$\\ \hline
        $s_{u}$\ vs.\ N&$4.059\cdot 10^{2}$&$-1.298\cdot 10^{-1}$&$0.938$\\ \hline
        $s_{u}$\ vs.\ $s_{f}$&$5.121\cdot 10^{2}$&$-1.329\cdot 10^{-1}$&$0.935$\\ \hline
        $\Delta\rho$\ vs.\ $\langle t \rangle$&$1.103\cdot 10^{16}$&$-9.975$&$0.949$\\ \hline
        $\Delta\rho$\ vs.\ N&$3.308\cdot 10^{-3}$&$1.110$&$0.991$\\ \hline
        $\langle t \rangle$\ vs.\ N&$7.061\cdot 10^{1}$&$-1.075\cdot 10^{-1}$&$0.970$\\ \hline
    \end{tabular}
    \vspace{5pt}
    \label{tab:fitTable}
    \caption{This table contains all the fits for the power-law graphs. The "a" and "b" values in each row follow the equation $y = ax^{b}$, and the $R^{2}$ is shown in the last column.}
\end{table}

\newpage

\section{ Discussion}

Hamilton's principle of stationary action has long been a cornerstone in physics, showing that the path taken by any system between two states is one that minimizes action for the most potentials in classical physics. In some cases, it is a saddle point never being a true maximum. Our research extends this principle to the realm of complex systems, proposing that the average action efficiency (AAE) serves as a predictor, measure, and driver of self-organization within these systems. By utilizing agent-based modeling (ABM), particularly through simulations of ant colonies, we demonstrate that systems naturally evolve towards states of higher organization and efficiency, consistent with the minimization of average physical action for one event in a system. In this simulation, as the number of agents in each run is fixed, all characteristics undergo a phase transition from an unorganized initial state to an organized final state. All of the characteristics are correlated with power-law relationships in the final state. This provides a new way of understanding self-organization and its driving mechanisms.

In an example of one agent, a state of the system where it has half of the action compared to another state, the system is calculated to have double the amount of organization. An extension of the model to open systems of $n$ agents provides a method for calculating the level of organization of any system. The significance of this result is that it provides a quantitative measure for comparing different levels of organization within the same system and for comparing different systems. 

The size-complexity rule can be summarized as the following: for a system to improve, it must become larger i.e. for a system to become more organized and action-efficient, it needs to expand. As a system's action efficiency increases, it can grow, creating a positive feedback loop where growth and action efficiency reinforce each other. The negative feedback loop is that the characteristics of a complex system cannot deviate much from the power law relationship. If we externally limit the growth of the system, we also limit the increase in its action efficiency. Then the action becomes stationary which means that the average action efficiency and the total action in the system stop increasing. Otherwise for unbounded, growing systems, action is dynamic, which means that the action efficiency and total action continue increasing. This applies to dynamic, open thermodynamic systems that operate outside of thermodynamic equilibrium and have flows of energy and matter. The growth of any system is driven by its increase in action efficiency. Without reaching a new level of action efficiency, growth is impossible. This principle is evident in both organisms and societies.

Other characteristics such as the total amount of action in the system, the number of events per unit of time, the internal entropy decrease, the density of agents, the amount of information in the system (measured in terms of pheromone levels), and the average time per event are strongly correlated and increase according to a power law function. Changing the population in the simulation influences all these characteristics through their power law relationships. Because these characteristics are interconnected, measuring one can provide the values of the others at any given time in the simulation using the coefficients in the power law fits (Table Appendix). If we consider the economy as a self-organizing complex system, we can find a logical explanation for the Jevons paradox, which may need to be renamed to Jevons rule, because it is an observation of a regular property of complex systems, and not an unexplained counter-intuitive fact, as it has been considered for a long time. 

We uncover a unit-total dualism, as in some of the characteristics, such as action and entropy, while the action and entropy per one event decrease, the total action, and entropy proportionally increase in the whole system. The unit and total quantities are correlated with power law equations. In the case of action, this leads to a dynamical action principle, where unit action is decreasing in self-organization, while total action is increasing. This variational principle is observed in self-organizing complex systems, and not in isolated agents. 

The formation of the path in the simulation is an emergent property in this system due to the interactions of the agents and is not specified in the rules of the simulation. This least action state of the system, which is its most organized state is predicted from the principle of least action. This means that we have a way to predict emergent properties in complex systems, using basic physics principles. The emergence of structure from the properties of the agents is a hallmark of self-organizing systems and it appears spontaneously. Emergence is a property of the entire system and not of its parts.

\section{ Conclusions}

This study reinforces the increase of average action efficiency during self-organization and with the size of systems as a driver and a measure for the evolution of complex systems. This offers new opportunities for understanding and describing the processes leading to increased organization in complex systems. It offers  prospects for future research, laying a foundation for more in-depth exploration into the dynamics of self-organization and potentially inspiring the development of new strategies for optimizing system performance and resilience.

Our findings suggest that self-organization is inherently driven by a positive feedback loop, where systems evolve towards states of minimal unit action and maximal organization. Self-organization driven by the action principles could be the simplest explanation and thus pass the Occam's razor. It could be the answer to "Why do complex systems self-organize at all?". Action efficiency always acts together with all other characteristics in the model, not in isolation. It drives self-organization through this mechanism of positive and negative feedback loops.

We found that this theory is working well for the current simulation. With additional details and features, it can be applied to more realistic systems. This model is testable and falsifiable. It needs to be always retested because every theory, every method, and every approach has its limits and needs to be extended, expanded, enriched, and detailed as new levels of knowledge are reached.  We expect this from all scientific theories and explanations. This model may be tested for any network, for example, metabolic networks, ecological networks, Internet, road networks, etc.

Our simulations demonstrate that the level of organization is inversely proportional to the average physical action required for system processes. This measure aligns with the principle of least action, a fundamental concept in physics, and extends its application to complex, non-equilibrium systems. The results from our ant colony simulations consistently show that systems with higher average action efficiency exhibit greater levels of organization, validating our hypothesis.

When the processes of self-organization are open-ended and continuous the stationary action principles do not apply anymore except in limited cases. We have dynamical action principles where the quantities are changing continuously, either increasing or decreasing.

We uncovered an extension of the principle of least action to complex systems, which can be an extended variational principle of decreasing unit action per one event in a self-organizing complex system, and it is connected with a power law relation to another mirror variational principle of increasing total action of the system. Other complexity variational principles are the decreasing unit entropy per one event in the system, and the increasing of the total entropy as the system grows, evolves, develops, and self-organizes. We term those polar sets of variational principles, unit-total duality. 

Other dualities to explore are, that the unit path curvature for one edge of the complex networks decreases, according to the Hertz's principle of least curvature, as the total curvature for traversing all paths in the system increases. The unit path constraint for the motion of one edge decreases, according to the Gauss principle of least constraint, as the total constraint for the motion of all agents as the system grows increases. There are possibly many more variational dualities to be uncovered in self-organizing, evolving, and developing complex systems. Those dualities can be used to analyze, understand, and predict the behavior of complex systems. This is one explanation for the size-complexity rule observed in nature and the scaling relationships in biology and society. The unit-total dualism is that as unit quantities decrease, with the system becoming more efficient as a result of self-organization, total quantities grow and both are connected with positive feedback and are correlated by a power law relation. As one example we find a logical explanation for the Jevons and other paradoxes, and the subsequent work of economists in this field, which are also unit-total dualities inherent to the functioning of self-organizing and growing complex systems. 

While our results are promising, our study has limitations. The simplified ant colony model used in our simulations does not capture the full spectrum of complexities and interactions present in real-world systems. Future research should aim to integrate more detailed and realistic models, incorporating environmental variability and agent heterogeneity, to test the universality and applicability of our findings more broadly and for specific systems.

Additionally, the interplay between average action efficiency and other organizational measures, such as entropy and order parameters, deserves further investigation. Understanding how these metrics interact could deepen our comprehension of complex system dynamics and provide a more holistic view of system organization.

The implications of our findings are significant for both theoretical research and practical applications. In natural sciences, this new measure can be used to quantify and compare the organization of different systems, providing insights into their evolutionary processes. In engineering and artificial systems, our model can guide the design of more efficient and resilient systems by emphasizing the importance of action efficiency. For example, in ecological and biological systems, understanding how organisms optimize their behaviors to achieve greater efficiency can inform conservation strategies and ecosystem management. In technology and artificial intelligence, designing algorithms and systems that follow the principle of least action can lead to more efficient processing and better performance. 

Our findings contribute to a deeper understanding of the mechanisms underlying self-organization and offer a novel, quantitative approach to measuring organization in complex systems. This research opens up exciting possibilities for further exploration and practical applications, enhancing our ability to design and manage complex systems across various domains. By providing a quantitative measure of organization that can be applied universally, we enhance our ability to design and manage complex systems across various domains. Future research can build on our findings to explore the dynamics of self-organization in greater detail, develop new optimization strategies, and create more efficient and resilient systems.
\newline

{\bf\large Acknowledgments}

The authors thank Assumption University for providing a creative atmosphere and funding and its Honors Program, specifically Prof. Colby Davie, for continuous research support and encouragement. Matthew Brouillet thanks his parents for their encouragement. Georgi Georgev thanks wife for patience and support for this manuscript.  
\newline

{\bf\large Author contributions }

Matthew J. Brouillet contributed all of the programming for simulation, data analysis and visualization, and formatting of data figures and tables. Georgi Yordanov Georgiev contributed the overall research direction, the questions for the study, the theory, model, methods, predictions, references, and the writing of most of the text. 

%\input{tex/Newmethods}
%\clearpage
\bibliography{main.bib}
%bib/New, bib/My, bib/EvolComplexity, bib/Information, bib/LAP, bib/MEPP, bib/Cybernetics, bib/Simulations, bib/SOC, bib/Math, bib/powerlaws}

\end{document}